\documentclass[fleqn,usenatbib]{mnras}

\usepackage{newtxtext,newtxmath}

\usepackage[T1]{fontenc}

\DeclareRobustCommand{\VAN}[3]{#2}
\let\VANthebibliography\thebibliography
\def\thebibliography{\DeclareRobustCommand{\VAN}[3]{##3}\VANthebibliography}

\usepackage{graphicx}	
\usepackage{amsmath}	
\usepackage{bm}
\usepackage{booktabs}
\usepackage{newtxtext,newtxmath}
\usepackage{threeparttable}

\usepackage[T1]{fontenc}
\usepackage{subcaption}

\graphicspath{{./}{figs/}}

\usepackage{scalerel,tikz}
\usetikzlibrary{svg.path}
\definecolor{orcidlogocol}{HTML}{A6CE39}
\tikzset{orcidlogo/.pic={\fill[orcidlogocol] svg{M256,128c0,70.7-57.3,128-128,128C57.3,256,0,198.7,0,128C0,57.3,57.3,0,128,0C198.7,0,256,57.3,256,128z}; \fill[white] svg{M86.3,186.2H70.9V79.1h15.4v48.4V186.2z} svg{M108.9,79.1h41.6c39.6,0,57,28.3,57,53.6c0,27.5-21.5,53.6-56.8,53.6h-41.8V79.1z M124.3,172.4h24.5c34.9,0,42.9-26.5,42.9-39.7c0-21.5-13.7-39.7-43.7-39.7h-23.7V172.4z} svg{M88.7,56.8c0,5.5-4.5,10.1-10.1,10.1c-5.6,0-10.1-4.6-10.1-10.1c0-5.6,4.5-10.1,10.1-10.1C84.2,46.7,88.7,51.3,88.7,56.8z};}}
\newcommand\orcidicon[1]{\href{https://orcid.org/#1}{\mbox{\scalerel*{
\begin{tikzpicture}[yscale=-1,transform shape]\pic{orcidlogo};
\end{tikzpicture}}{|}}}}

\title[Weakly-collisional and MHD turbulent dynamo]{A comparison of the turbulent dynamo in weakly-collisional and collisional plasmas: from subsonic to supersonic turbulence}
\author[Achikanath Chirakkara et al.]{
Radhika Achikanath Chirakkara,$^{\orcidicon{0000-0001-5583-5038}\,1}$\thanks{E-mail: \href{mailto:radhika.achikanathchirakkara@anu.edu.au}{radhika.achikanathchirakkara@anu.edu.au}}
Christoph Federrath,$^{\orcidicon{0000-0002-0706-2306}\,1,2}$
and Amit Seta$^{\orcidicon{0000-0001-9708-0286}\,1}$
\\
$^{1}$Research School of Astronomy and Astrophysics, Australian National University, Canberra, ACT 2611, Australia\\
$^{2}$Australian Research Council Centre of Excellence in All Sky Astrophysics (ASTRO3D), Canberra, ACT 2611, Australia
}
\date{Accepted XXX. Received YYY; in original form ZZZ}

\pubyear{2024}
\newcommand\Eq[1]{Eq.\,\ref{#1}}
\newcommand\Fig[1]{Fig.~\ref{#1}}
\newcommand\Sec[1]{Sec.~\ref{#1}}
\newcommand\Tab[1]{Tab.~\ref{#1}}
\newcommand\App[1]{Appendix~\ref{#1}}

\newcommand\plus{\texttt{+}}
\newcommand\minus{\texttt{-}}

\newcommand{\Mach}{{\mathcal{M}}}      
\newcommand{\Pm}{{\rm Pm}}
\renewcommand{\Re}{\text{Re}} 
\newcommand{\Rm}{{\rm Rm}}

\renewcommand{\vec}[1]{\boldsymbol{#1}}	
\newcommand{\dd}{\mathrm{d}}        

\newcommand{\cm}{{\rm cm}}    
\newcommand{\km}{{\rm km}}    
\newcommand{\Mpc}{{\rm Mpc}}  

\newcommand{\s}{{\rm s}}      
\newcommand{\kms}{\km\s^{-1}}    

\newcommand{\G}{{\rm G}}      

\newcommand{\TeV}{{\rm TeV}}  

\newcommand{\Emag}{E_{\rm{m}}/E_{\rm{m0}}}

\newcommand{\initmagnetisation}{(r_{\rm Larmor}/L)_{0}}
\newcommand{\Einit}{(E_{\rm{mag}}/E_{\rm{kin}})_{0}}
\newcommand{\betainit}{\beta_{0}}
\newcommand{\ngrid}{N_{\rm{grid}}}
\newcommand{\nppc}{N_{\rm{ppc}}}
\newcommand{\mfp}{\lambda_{\rm mfp}}
\newcommand{\vel}{\vec{v}}
\newcommand{\pos}{\vec{r}}
\renewcommand{\vec}[1]{\mathbf{#1}}
\newcommand{\Vturb}{u_{\rm turb}}
\newcommand{\Vtherm}{u_{\rm th}}
\newcommand{\Lturb}{L_{\rm turb}}
\newcommand{\tth}{t_{\rm th}}
\newcommand{\tcool}{t_{\rm cool}}
\newcommand{\ted}{t_{0}}

\newcommand{\E}{\vec{E}}
\newcommand{\B}{\vec{B}}
\newcommand{\J}{\vec{J}}

\newcommand{\dl}{\Delta l}

\newcommand{\ahkash}{\texttt{AHKASH} \,}

\newcommand{\bcdot}{\,\mathbf{\cdot}\,}
\newcommand{\grad}{\mathbf{\nabla}}
\newcommand{\btimes}{\,\mathbf{\times}\,}
\newcommand{\Pkin}{P_{\mathrm{kin}}}

\begin{document}
\label{firstpage}
\pagerange{\pageref{firstpage}--\pageref{lastpage}}
\maketitle

\begin{abstract}
Weakly-collisional plasmas, such as the solar wind or the intra-cluster medium (ICM) of galaxy clusters, evolve in the presence of dynamically strong magnetic fields. The turbulent dynamo can amplify magnetic fields to such levels by converting turbulent kinetic energy into magnetic energy. While extensively studied in collisional magnetohydrodynamic (MHD) simulations, the weakly-collisional regime has only been explored recently. Here, we determine the properties of the weakly-collisional turbulent dynamo in the exponential ``kinematic" growth phase in both the subsonic and the previously unexplored supersonic regime of turbulence, using hybrid particle-in-cell (HPIC) and MHD simulations. We conduct a large parameter study, fixing the magnetic Reynolds number, $\Rm = 500$, and the initial ratio of the magnetic to kinetic energy, $\Einit = 10^{-10}$, and then vary the kinetic Reynolds number, $\Re = 500$, $50$, and $5$, for the MHD simulations. In the HPIC runs, only $\Rm=500$ is controlled, while $\Re$ emerges self-consistently from wave-particle interactions. We find that the velocity and magnetic field structures, probability distribution functions, and power spectra of the HPIC runs are similar to that of the MHD dynamo with $\Re \sim 50-500$ and $\Re \sim 500$ in the subsonic and supersonic regimes, respectively. Using MHD scaling relations, we infer $\Re_{\rm inferred}=480^{+170}_{-250}$ and $690^{+360}_{-360}$ in the subsonic and supersonic weakly-collisional plasma, respectively. Overall, we find that the turbulent dynamo shares similar physical properties in both weakly-collisional and collisional plasmas. Our results of the weakly-collisional turbulent dynamo may have relevant applications to the solar wind, weakly-collisional shocks, and the hot ICM.
\end{abstract}



\section{Introduction}
Plasmas can be characterised collisional or weakly=collisional based on the ratio of the mean free path, $\mfp$, which is the average distance that charged particles travel before interacting with another particle, and the characteristic system size, $L$. The system is considered collisional if the mean free path is much smaller than the system size, as is the case in the interstellar medium, where $\mfp/L \ll 1$ from the hot ionised medium to the cold molecular phase \citep{Ferriere2020}.
Such plasmas have been conventionally studied using collisional magnetohydrodynamical (MHD) simulations. However, for many plasmas such as the hot intracluster medium (hot ICM) of galaxy clusters, $\mfp/L \sim 0.1 - 1$, making this plasma weakly-collisional \citep{Schekochihin&Cowley2006}. In this weakly-collisional regime, the MHD equations are inadequate for modelling the plasma, and we have to resort to kinetic methods. This is also true for the solar wind, cosmic-ray acceleration, and any scenario involving collisionless shocks, kinetic instabilities, and magnetic reconnection \citep{Kulsrud2005}. This study aims to understand the amplification of magnetic fields in such weakly-collisional turbulent plasmas, particularly in the previously unexplored compressible and supersonic regime, and to contrast it with the well-studied MHD regime. Compressibility is relevant for many of these systems, and here we consider the ICM of galaxy clusters as a specific example of a compressible weakly- collisional plasma \citep{Hoang+2017}.

The hot ICM has extremely high temperatures of $\approx 10^{7} - 10^{8}$~K, and is very diffuse with particle number densities of $\approx 10^{-3} \, \cm^{-3}$ \citep{Simionescu+2019, Kunz+2022}. Dynamically strong magnetic fields have been observed in the hot ICM with $\sim \mu \G$ strength up to $\Mpc$ scales \citep{Bonafedeetal2010, Bonafede+2013,Botteon+2022}. From gamma-ray observations of $\TeV$ blazars, a lower limit on the magnetic field strength of $\sim 10^{-12} - 10^{-10} \, \mu \G$ in the voids of the large-scale structure of the Universe has been inferred \citep{Neronov2010, Tavecchioetal2010MNRAS,TaylorVovkNeronov2011, Vovketal2012ApJ, Finketal2015, FermiLATcollaboration2018}. These magnetic fields, which can be of astrophysical or primordial origin, are much smaller than the $\mu \G$ magnetic fields observed in the hot ICM. Thus, there must be a powerful amplification mechanism to explain the fields observed today: the turbulent dynamo, which is the process of converting the kinetic energy of turbulence to magnetic energy \citep{Kazantsev1968, Moffatt1978, B&S2005}.

However, the turbulent dynamo process requires sustained turbulence. Galaxy mergers, wakes of infalling galaxies, and AGN can indeed drive the required turbulence in the hot ICM of galaxy clusters \citep{Simionescu+2019}. Turbulent motions $\approx 160 \, \kms$ have been observed in the Perseus cluster by the Hitomi collaboration \citep{HitomiCollaboration2016}. Recent X-ray spectroscopy studies have inferred subsonic Mach numbers, $\Mach \sim 0.2 - 0.4$ in galaxy clusters, reaching transonic levels ($\Mach \sim 1$) close to the cluster centres \citep{Gatuzz+2022a, Gatuzz+2022b, Gatuzz+2023}. Additionally, high-fidelity radio observations reveal supersonic Mach numbers, $\Mach \sim 2.5$, in shocked regions of the sausage relic \citep{Hoang+2017}. Thus, subsonic, incompressible theories and simulations \citep{Rinconetal2016,St-Onge&Kunz2018,AchikanathChirakkara+2023} are likely insufficient to explain the dynamo amplification in these cosmic structures.

The turbulent dynamo, broadly, has three stages. First, a kinematic regime (1), where the weak seed magnetic field grows exponentially on small length scales; then a transition regime (2), where the magnetic field grows in a power-law fashion, often described as linear for subsonic flows \citep{Cho+2009, Xu&Lazarian2016, Seta&Federrath2020, Beattie+2023}, and the coherence length scale of the magnetic field increases; and finally, a saturation phase (3), where the fields cease growing, but are still maintained by the dynamo. The turbulent dynamo has been widely studied using collisional MHD simulations \citep{Schekochihin+2004, Haugen+2004a, Federrathetal2014ApJ, Federrath2016, Seta+2020, Seta&Federrath2020, Seta&Federrath2021b, Seta&Federrath2021a, Seta&Federrath2022, KrielEtAl2022, Gent+2023}, and its role in amplifying weak seed magnetic fields to strong dynamical levels have been explored in many astrophysical environments including star-forming molecular clouds \citep{Sur+2010, Federrath+2011}, young galaxies \citep{Schoberetal2013}, and galaxy clusters \citep{Subramanian+2006,Bhat&Subramanian2013, Vazza+2018, Sur+2019}. Properties of the MHD turbulent dynamo, like the growth rate of magnetic energy and the saturation efficiency, depend on the Mach number of the plasma and the nature of the turbulent driving \citep{CFetal11, AchikanathEtAl2021}. Recent laboratory experiments have also demonstrated the turbulent dynamo mechanism in different Mach number regimes \citep{Tzeferacos+2017, Tzeferacos+2018, Chen+2020, Bott+2021a, Bott+2021b}, confirmed in simulations \citep{HuEtAl2022,HewFederrath2023}. The plasmas in the above numerical studies and experiments are collisional \citep[e.g.~$\mfp/L \sim 10^{-3}$, see][]{Tzeferacos+2018}.

The possibility of a turbulent dynamo in weakly-collisional plasmas has only recently been explored, with several approaches employed to study the weakly-collisional turbulent dynamo \citep{Rinconetal2016, St-Onge&Kunz2018, St-Ongeetal2020, AchikanathChirakkara+2023, AchikanathChirakkara+2024, Sironi+2023, Zhou+2024}. For example, \citet{Sironi+2023} and \citet{Zhou+2024} use the fully kinetic particle-in-cell (PIC) approach, which models all plasma components, including ions and electrons, as particles. This makes the approach computationally demanding. In the hybrid-kinetic approach, ions are modelled kinetically, while electrons are treated as a fluid. \citet{St-Onge&Kunz2018} and \citet{AchikanathChirakkara+2023} use the hybrid particle-in-cell (hybrid PIC) approach, where Lagrangian particles sample the moments of the ion distribution function, while electrons are described as a fluid. The hybrid PIC approach is computationally less expensive and can be used to study the system on ion length- and time-scales, which are much larger than the electron length- and time-scales. Therefore, we use the hybrid PIC approach in the present work.

The properties of the weakly-collisional turbulent dynamo for different parameter regimes, especially in the supersonic regime, are still unknown. Furthermore, it is not clear how the properties of the weakly-collisional turbulent dynamo compare to those of its well-studied collisional MHD counterpart. In this paper, we study the weakly-collisional turbulent dynamo using hybrid PIC simulations in the subsonic and the previously unexplored supersonic regime and compare it with the MHD turbulent dynamo. 

Unlike in MHD, where viscosity is a parameter that can be set in numerical simulations, the effective viscosity of a weakly-collisional plasma is determined self-consistently by wave-particle interactions \citep{Schekochihin&Cowley2006}. This makes it challenging to ascertain the kinetic Reynolds number, viscous dissipation scale, and magnetic Prandtl number -- all critical plasma parameters for studying the turbulent dynamo in weakly-collisional plasmas. In this work, we address this by applying viscous and magnetic scaling relations from studies of the MHD turbulent dynamo to the weakly-collisional regime.

The rest of this work is organised in the following way. We describe the hybrid-kinetic and MHD equations, the numerical methods, and the initial conditions and parameters for the turbulent dynamo simulations in \Sec{sec:methods}. In \Sec{sec:results}, we discuss the morphology, time evolution, probability density functions, and the power spectra for the weakly-collisional and the collisional MHD turbulent dynamo in the subsonic and the supersonic regime. The kinetic Reynolds number and the turbulent dissipation scale of the weakly-collisional turbulent dynamo simulations are estimated using MHD scaling relations in \Sec{sec:Re_scaling}. Finally, we conclude our study in \Sec{sec:conclusions}.

\section{Methods}
\label{sec:methods}
We use the hybrid PIC method (henceforth referred to as HPIC) \ahkash \citep{AchikanathChirakkara+2024} to perform numerical experiments of the weakly-collisional dynamo, while the collisional MHD simulations presented in this study are performed with the HLL5R Riemann scheme \citep{Waagan2011}. Both HPIC and MHD methods used here are implemented in the \texttt{FLASH} code framework \citep{Fryxelletal2000, Dubey+2008a, Dubey+2008b}.

\subsection{Hybrid-kinetic equations} \label{sec:hk_eqns}
For the simulations of weakly-collisional plasmas, we solve the hybrid-kinetic equations (written here in SI units) using the HPIC method. In the HPIC approach, the positively charged ions are treated as particles, and the positions and velocities of the particles are evolved using the equations of motion,
\begin{align}
        & \frac{\dd \pos_{p}} {\dd t} = \vel_{p},
        \label{eqn:part_evolution1} \\
        & \frac{\dd \vel_{p} }{\dd t} = \frac{q_{\rm i}}{m_{\rm i}}\left(\E + \vel_{p} \times \B\right) + \vec{f},  \label{eqn:part_evolution2} 
\end{align}
where $\pos_{p}$ and $\vel_{p}$ are the position and velocity of the $p^{\rm th}$ particle, respectively, with ${p} = 1, 2, \cdots, {P}$, where $P$ is the total number of particles. The quantities $q_{\rm i}$ and $m_{\rm i}$ are the ion charge and mass, respectively. The ions (particles) move under the influence of the electric ($\E$) and magnetic ($\B$) fields and sample the moments of the ion distribution function. The additional turbulent acceleration field, $\vec{f}$, further described in \Sec{sec:turb_driving}, is used to drive turbulence in the plasma. 

The electrons are treated as a massless fluid in the hybrid-kinetic approach. The electric field is calculated using the following Ohm's law \citep{Rosin+2011},
\begin{equation}
\E = \frac{\left( \J - \J_{\rm i} \right) \btimes \B}{\rho_{\rm i}} - \frac{\grad p_{\rm e}}{\rho_{\rm i}}+ \mu_{0} \eta \J- \mu_{0} \eta_{\rm hyper} \nabla^{2} \J,
\label{eqn:ohms_law}
\end{equation}
where $\J = (\nabla \times \B)/\mu_{0}$ is the total current, i.e., the sum of the ion and electron currents. The charge density, $\rho_{\rm i}$, and ion current, $\J_{\rm i}$, are sampled from the particles. The electron pressure, $p_{\rm e}$, is modelled using the ideal gas equation of state, $p_{\rm e} = n_{\rm e} k_{\rm B} T_{\rm e}$, where $n_{\rm e}$, $T_{\rm e}$, and $k_{\rm B}$ are the electron number density, electron temperature, and the Boltzmann constant, respectively. We assume the electrons are isothermal ($T_{\rm e} = {\rm constant}$). We also assume that the plasma is quasi-neutral, $\rho_{\rm e} = - \rho_{\rm i}$, where $\rho_{\rm e}$ is the electron charge density. $\mu_0$ is the magnetic permeability constant. The diffusivity, $\eta$, sets the Ohmic dissipation of the plasma. The final term on the right-hand side of \Eq{eqn:ohms_law} is the numerical hyper-resistivity, and $\eta_{\rm hyper}$ is the hyper-diffusivity co-efficient, which is used to suppress numerical fluctuations near the grid scale.

The magnetic field is computed using Faraday's law,
\begin{equation}
    \frac{\partial \B}{\partial t} = - \nabla \times \E,
    \label{eqn:faradays_law}  
\end{equation}
solved using the constrained transport method, which ensures that the magnetic fields are divergence-free ($\nabla \cdot \B = 0$). We use the cloud-in-cell kernel to perform interpolation operations between the particles and the computational grid to compute moments of the ion distribution. We perform a smoothing operation on the moments sampled from particles, to reduce particle noise due to limited sampling of the ion distribution function (e.g.~see Fig.~12 in \cite{AchikanathChirakkara+2024} for further details). The hybrid PIC simulations in this work are done using the \ahkash code, which is described in detail with relevant tests in \citet{AchikanathChirakkara+2024}.

\subsection{Cooling for hybrid particle-in-cell simulations}
Various physical processes can act to heat and cool a plasma. A balance between heating and cooling leads to a constant plasma temperature, and thus a constant thermal speed of the ions in many physical systems. Therefore, it is often desirable to model a plasma with constant temperature. In hybrid PIC, one can directly set an isothermal equation of state for the electrons, as discussed in \Sec{sec:hk_eqns}. However, this approach is not applicable to ions, which are treated as particles. Therefore, a cooling method is necessary to keep the ions isothermal and ensure they maintain the same temperature as the electrons. Without a cooling method, various heating sources, such as turbulence or shock dissipation, make it impossible to achieve steady-state turbulence with a time-independent root mean square (rms) Mach number. 

To mitigate this issue, we have introduced a cooling method based on locally rescaling the ion thermal speed, $\Vtherm$ \citep[Sec.~2.11 and Sec.~5 in][for the description of the method and related tests]{AchikanathChirakkara+2024}. This cooling method maintains isothermal conditions locally and globally throughout the computational domain and in time. In the subsonic regime, cooling is performed on a cooling timescale, $(\Delta t)_{\rm cool} = 0.1 \tcool$ and $\tcool = \tth/\Mach$, where $\tth = \dl/\Vtherm$ is the thermal crossing time (the time taken by sound waves to travel across a grid cell of size $\dl$) and $\Mach$ is the Mach number of the plasma. In the supersonic regime, a shorter cooling timescale, $(\Delta t)_{\rm cool} = 0.01 \tcool$, is required as more energy compared to the thermal energy is injected into the plasma in the supersonic case and dissipative shocks are present.

\subsection{Magnetohydrodynamic equations}
For the MHD simulations, we use the HLL5R Riemann scheme \citep{bouchut2007multiwave, bouchut2010multiwave, Waagan2011}, solving the isothermal compressible MHD equations (written in CGS units),
\begin{align}
        & \frac{\partial \rho}{\partial t} + \nabla \cdot (\rho \vec{u}) = 0,
        \label{eqn:continuity_eqn} \\
        & \frac{\partial (\rho \vec{u})}{\partial t} + \nabla \cdot (\rho \vec{u} \otimes \vec{u} - \B \otimes \B) = -\nabla p + \nabla \cdot (2 \nu \rho S) + \rho \vec{f},  
        \label{eqn:navier_stokes} \\
        & \frac{\partial \B}{\partial t} = \nabla \times (\vec{u} \times \B) + \eta  \nabla^{2} \B,
        \label{eqn:induction_eqn} \\
        & \nabla \cdot \B = 0,
\end{align}
where $\rho$ is the mass density of the gas, $\vec{u}$ is the velocity, $\B$ is the magnetic field, $p$ is the pressure, which includes both the thermal pressure, $p_{\rm th}$ and the magnetic pressure, $p_{\rm mag}=B^2/(8\pi)$, and $\vec{f}$ is the turbulent acceleration field, same as used in \Sec{eqn:part_evolution2}. The viscosity and Ohmic diffusivity are denoted by $\nu$ and $\eta$, respectively. The rate of strain tensor, $S_{ij}$, is given by $S_{ij} = 1/2 (\partial_i u_j + \partial_j u_i - (2/3) \delta_{ij} \nabla \cdot \vec{u})$. The MHD equations are closed with an isothermal equation of state, $p_{\rm th} = \rho \Vtherm^{2}$, with a constant sound speed $c_{\rm s}$. So, the energy equation of MHD does not need to be solved. We set the viscosity and diffusivity such that specific target Reynolds numbers are achieved given the available numerical resolution \citep{Malvadi&Federrath2023} in our numerical experiments (detailed in Sec.~\ref{sec:ics}).

\subsection{Turbulent driving}
\label{sec:turb_driving}
The turbulent driving field, $\vec{f}$, is modelled using the Ornstein--Uhlenbeck process using the turbulence generator \texttt{TurbGen} \citep{Federrath+2010,FederrathEtAl2022ascl} for both the HPIC and MHD cases. Turbulence is driven on large scales, i.e., wave numbers satisfying $kL/2\pi =(1,3)$, where $L$ is the side length of the cubic computational domain. The driving amplitude is controlled by a parabolic function that peaks at $k_{\rm turb}L/2\pi = 2$, and falls to zero at $kL/2\pi = 1, 3$. In this study, we focus on purely solenoidal driving $(\grad \bcdot \vec{f} = 0)$, which injects solenoidal acceleration modes into the plasma with exactly the same $f$ for the HPIC and MHD simulations.

\subsection{Plasma parameters and initial conditions} \label{sec:ics}
This section discusses the important parameters and initial conditions of our HPIC and MHD turbulent dynamo numerical simulations. We perform the HPIC simulations on a periodic computational box with $\ngrid^{3} = 128^{3}$ grid cells and $\nppc = 100$ particles per cell. We also test and demonstrate numerical convergence by varying the grid resolution, $\ngrid^{3} = 64^{3}$ and $256^{3}$, and the particle resolution, $\nppc = 50$ and $200$ in \App{app:hybridPIC_conv}. For a better comparison, the MHD simulations are also performed on $\ngrid^{3} = 128^{3}$ grid cells with periodic boundary conditions. To demonstrate the convergence of the MHD simulations, we vary the grid resolution, $\ngrid^3 = 64^3$ and $256^3$ in \App{app:MHD_conv}. 

\begin{table*}
\centering
\caption{List of simulations with the corresponding model name, grid resolution ($\ngrid^{3}$), particle-per-cell count ($\nppc$), Mach number ($\mathcal{M}$), initial magnetic to kinetic energy ratio ($\Einit$), initial plasma beta ($\betainit$), initial Larmor ratio ($\initmagnetisation$), magnetic Reynolds number ($\Rm$), kinetic Reynolds number ($\Re$), magnetic Prandtl number ($\Pm$), dynamo growth rate of the magnetic energy ($\Gamma$), and the magnetic dissipation scale ($k_{\eta})$. }
\setlength{\tabcolsep}{2.5pt}
\begin{tabular}{llccccccccccc}
\hline
Ser. No & Model & $\ngrid^{3}$ & $\nppc$ & $\mathcal{M}$ & $\Einit$ & $\betainit$ & $\initmagnetisation$ & $\Rm$ & $\Re$ &  $\Pm$ & $\Gamma$ ($t_{0}^{-1}$) & $k_{\eta}$\\
\hline
1 & \texttt{HPICM0.2Rm500} & $128^{3}$ & 100 & 0.20$\pm$0.02 & $10^{-10}$ & $5 \times 10^{11}$  & $10^{2}$ & 502$\pm$41 & $-$ & $-$ & 0.49$\pm$0.05 & 9.5$\pm$1.0\\
2 & \texttt{MHDM0.2Rm500Re500} & $128^{3}$ & $-$ & 0.20$\pm$0.01 & $10^{-10}$ & $5 \times 10^{11}$ & $-$ & 500$\pm$26 & 500$\pm$26 & 1 & 0.38$\pm$0.02 & 10.8$\pm$1.0\\
3 & \texttt{MHDM0.2Rm500Re50} & $128^{3}$ & $-$ & 0.20$\pm$0.01 & $10^{-10}$ & $5 \times 10^{11}$ & $-$ & 504$\pm$35 & 50.4$\pm$3.5 & 10 & 0.55$\pm$0.01 & 8.4$\pm$1.0\\
4 & \texttt{MHDM0.2Rm500Re5} & $128^{3}$ & $-$ & 0.20$\pm$0.02 & $10^{-10}$ & $5 \times 10^{11}$ & $-$ & 500$\pm$40 & 5.0$\pm$0.4 & 100 & 0.54$\pm$0.03 & 7.2$\pm$1.0\\
5 & \texttt{HPICM2Rm500} & $128^{3}$ & 100 & 1.91$\pm$0.12 & $10^{-10}$ & $5 \times 10^{9}$ & $10^{2}$ & 476$\pm$30 & $-$ & $-$ & 0.37$\pm$0.05 & 9.9$\pm$1.1\\
6 & \texttt{MHDM2Rm500Re500} & $128^{3}$ & $-$ & 1.95$\pm$0.09 & $10^{-10}$ & $5 \times 10^{9}$ & $-$ & 488$\pm$23 & 488$\pm$23 & 1 & 0.14$\pm$0.05 & 9.3$\pm$1.0\\
7 & \texttt{MHDM2Rm500Re50} & $128^{3}$ & $-$ & 1.99$\pm$0.10 & $10^{-10}$ & $5 \times 10^{9}$ & $-$ & 497$\pm$25 & 49.7$\pm$2.5 & 10 & 0.44$\pm$0.03 & 8.1$\pm$1.0\\
8 & \texttt{MHDM2Rm500Re5} & $128^{3}$ & $-$ & 1.92$\pm$0.14 & $10^{-10}$ & $5 \times 10^{9}$ & $-$ & 479$\pm$36 & 4.8$\pm$0.4 & 100 & 0.55$\pm$0.04 & 7.1$\pm$1.0\\
 \hline
9 & \texttt{HPICM0.2Rm500Nppc50} & $128^{3}$ & 50 & 0.20$\pm$0.02 & $10^{-10}$ & $5 \times 10^{11}$  & $10^{2}$ & 501$\pm$42 & $-$ & $-$ & 0.51$\pm$0.04 & 9.5$\pm$1.1\\
10 & \texttt{HPICM0.2Rm500Nppc200} & $128^{3}$ & 200 & 0.20$\pm$0.02 & $10^{-10}$ & $5 \times 10^{11}$  & $10^{2}$ & 503$\pm$41 & $-$ & $-$ & 0.47$\pm$0.06 & 9.5$\pm$1.0\\
11 & \texttt{HPICM2Rm500Nppc50} & $128^{3}$ & 50 & 1.91$\pm$0.11 & $10^{-10}$ & $5 \times 10^{9}$ & $10^{2}$ & 478$\pm$27 & $-$ & $-$ & 0.41$\pm$0.01 & 10.1$\pm$1.1\\
12 & \texttt{HPICM2Rm500Nppc200} & $128^{3}$ & 200 & 1.92$\pm$0.11 & $10^{-10}$ & $5 \times 10^{9}$ & $10^{2}$ & 479$\pm$28 & $-$ & $-$ & 0.39$\pm$0.01 & 10.0$\pm$1.2\\
\hline
13 & \texttt{HPICM0.2Rm500Ngrid64} & $64^{3}$ & 100 & 0.20$\pm$0.02 & $10^{-10}$ & $5 \times 10^{11}$  & $10^{2}$ & 499$\pm$46 & $-$ & $-$ & 0.56$\pm$0.02 & 8.9$\pm$1.0\\
14 & \texttt{HPICM0.2Rm500Ngrid256} & $256^{3}$ & 100 & 0.20$\pm$0.02 & $10^{-10}$ & $5 \times 10^{11}$  & $10^{2}$ & 500$\pm$40 & $-$ & $-$ & 0.48$\pm$0.06 & 9.4$\pm$1.0\\
15 & \texttt{HPICM2Rm500Ngrid64} & $64^{3}$ & 100 & 1.85$\pm$0.10 & $10^{-10}$ & $5 \times 10^{9}$ & $10^{2}$ & 463$\pm$26 & $-$ & $-$ & 0.44$\pm$0.05 & 8.8$\pm$1.0\\
16 & \texttt{HPICM2Rm500Ngrid256} & $256^{3}$ & 100 & 1.98$\pm$0.12 & $10^{-10}$ & $5 \times 10^{9}$ & $10^{2}$ & 494$\pm$29 & $-$ & $-$ & 0.46$\pm$0.05 & 10.7$\pm$1.4\\
\hline
17 & \texttt{MHDM0.2Rm500Re500Ngrid64} & $64^{3}$ & $-$ & 0.20$\pm$0.01 & $10^{-10}$ & $5 \times 10^{11}$ & $-$ & 500$\pm$26 & 500$\pm$26 & 1 & 0.43$\pm$0.02 & 7.7$\pm$1.0\\
18 & \texttt{MHDM0.2Rm500Re500Ngrid256} & $256^{3}$ & $-$ & 0.20$\pm$0.01 & $10^{-10}$ & $5 \times 10^{11}$ & $-$ & 503$\pm$34 & 503$\pm$34 & 1 & 0.45$\pm$0.02 & 12.6$\pm$1.0\\
19 & \texttt{MHDM2Rm500Re500Ngrid64} & $64^{3}$ & $-$ & 1.93$\pm$0.09 & $10^{-10}$ & $5 \times 10^{9}$ & $-$ & 483$\pm$22 & 483$\pm$22 & 1 & 0.06$\pm$0.01 & 11.2$\pm$5.2\\
20 & \texttt{MHDM2Rm500Re500Ngrid256} & $256^{3}$ & $-$ & 1.96$\pm$0.09 & $10^{-10}$ & $5 \times 10^{9}$ & $-$ & 489$\pm$22 & 489$\pm$22 & 1 & 0.16$\pm$0.04 & 11.1$\pm$1.0\\
\hline
\end{tabular}
\label{table:sims}
\end{table*}

\subsubsection{Plasma parameters}
First, the plasma parameters that are important for the turbulent dynamo problem are discussed. The Mach number, $\Mach$, quantifies the compressibility of the plasma and is defined as 
\begin{equation}
    \Mach = \frac{\Vturb}{\Vtherm},
\end{equation}
where $\Vturb$ is the turbulent speed and $\Vtherm$ is the thermal speed. We study the turbulent dynamo in subsonic and supersonic regimes with $\Mach = 0.2$ and $\Mach = 2$, respectively. We fix the thermal speed while varying the turbulent speed to achieve different Mach numbers. The turbulent crossing time, $\ted = \Lturb/\Vturb$, where $\Lturb = L/2$. The magnetic Reynolds number is defined as 
\begin{equation}
    \Rm = \frac{\Vturb \Lturb}{\eta},
\end{equation}
where $\eta$ is the Ohmic diffusivity. We set $\Rm \approx 500$ in all our simulations, which is resolvable with a grid resolution of $\ngrid^{3} = 128^{3}$ in both the subsonic and supersonic regime (See Fig.~ 8 of \citet{Malvadi&Federrath2023}). We also use numerical hyper-resistivity in the HPIC simulations to remove grid-scale fluctuations in the magnetic field. The hyper-resistive Reynolds number is defined as 
\begin{equation}
    \Rm_{\rm hyper} = \frac{\Vturb \Lturb^{3}}{\eta_{\rm hyper}}.
\end{equation}
We set $\Rm_{\rm hyper} = \ngrid^{10/3}$ for the HPIC simulations, where $\ngrid$ is the grid resolution (Sec.~ 2.5.2 of \citet{AchikanathChirakkara+2023}). For the MHD simulations, we set the kinetic Reynolds number,
\begin{equation}
    \Re = \frac{\Vturb \Lturb}{\nu},
\end{equation}
to $\Re = 500, 50$, and $5$ in different simulations, where $\nu$ is the kinematic viscosity. The simulations with $\Re = 50$ and $\Re=5$ are highly viscous and not turbulent. While $\Re$ is controlled directly in the MHD simulations, in HPIC, the viscosity of the plasma arises self-consistently from wave-particle interactions.

The magnetic Prandtl number is defined as
\begin{equation} \label{eqn:Pm}
    \Pm = \frac{\Rm}{\Re}.
\end{equation}
Changing $\Re$ in the MHD simulations varies $\Pm = 1, 10$ and 100 for $\Re = 500, 50$ and 5, respectively. The $\Re$ and, consequently, the $\Pm$ of the weakly-collisional turbulent dynamo cannot be set as for the MHD simulations and we aim to infer the self-consistently developed $\Re$ and $\Pm$ from our HPIC simulations.

\subsubsection{Initial conditions}
Here, we discuss the initial conditions for the turbulent dynamo simulations. The initial magnetic to turbulent kinetic energy ratio, $\Einit$, can be re-written in the form of the initial plasma beta, $\betainit$, and the Mach number, as
\begin{equation}
    \Einit = \frac{2}{\betainit \Mach^{2}},
\end{equation}
where plasma $\beta$ is the ratio of thermal to magnetic pressure. An initial uniform magnetic field is set up for both the HPIC and MHD runs. We set $\Einit = 10^{-10}$ in all simulations to initialise an extremely weak magnetic field that can be amplified by the turbulent dynamo.

The Larmor ratio, defined as the ratio of the Larmor radius to the simulation box size, is an important parameter for studying the weakly-collisional turbulent dynamo and quantifies the magnetisation level of the plasma \citep{AchikanathChirakkara+2023}. The initial Larmor ratio is
\begin{equation}
    \initmagnetisation = \frac{m_{\rm i} \Vtherm}{q_{\rm i} b_{0} L},
\end{equation}
where $b_{0}$ is the initial magnetic field strength. We set $\initmagnetisation = 100$ in the HPIC simulations. \citet[][sec~2.5]{AchikanathChirakkara+2023} provide further details on the initial conditions used here, in particular on how they relate various plasma parameters. The HPIC simulations are performed using the hybrid-precision method, which significantly reduces the computational cost of these runs \citep{Federrath+2021,AchikanathChirakkara+2024}. 

\subsubsection{Simulation table}
For all simulations presented in this work, we tabulate the grid and particle resolutions, initial conditions, the magnetic and kinetic Reynolds numbers, the magnetic Prandtl number, the measured dynamo growth rate, and the magnetic energy dissipation scale in \Tab{table:sims}.

\section{Results}
\label{sec:results}
In this section, we discuss the weakly-collisional HPIC and collisional MHD turbulent dynamo in both the subsonic and supersonic regimes. First, we study the density, velocity, and magnetic field structures in \Sec{sec:slice_plots}. The time evolution and growth rate of the dynamo in the kinematic regime are discussed in \Sec{sec:time_evol}. The probability density functions of the density, velocity, and magnetic field are studied in \Sec{sec:pdfs}. Finally, we discuss the power spectra of the magnetic energy, current, and turbulent kinetic energy in \Sec{sec:spectra}.

\subsection{Density, velocity, and magnetic structures} \label{sec:slice_plots}
\begin{figure*}
    \centering
    \includegraphics[scale=0.88]{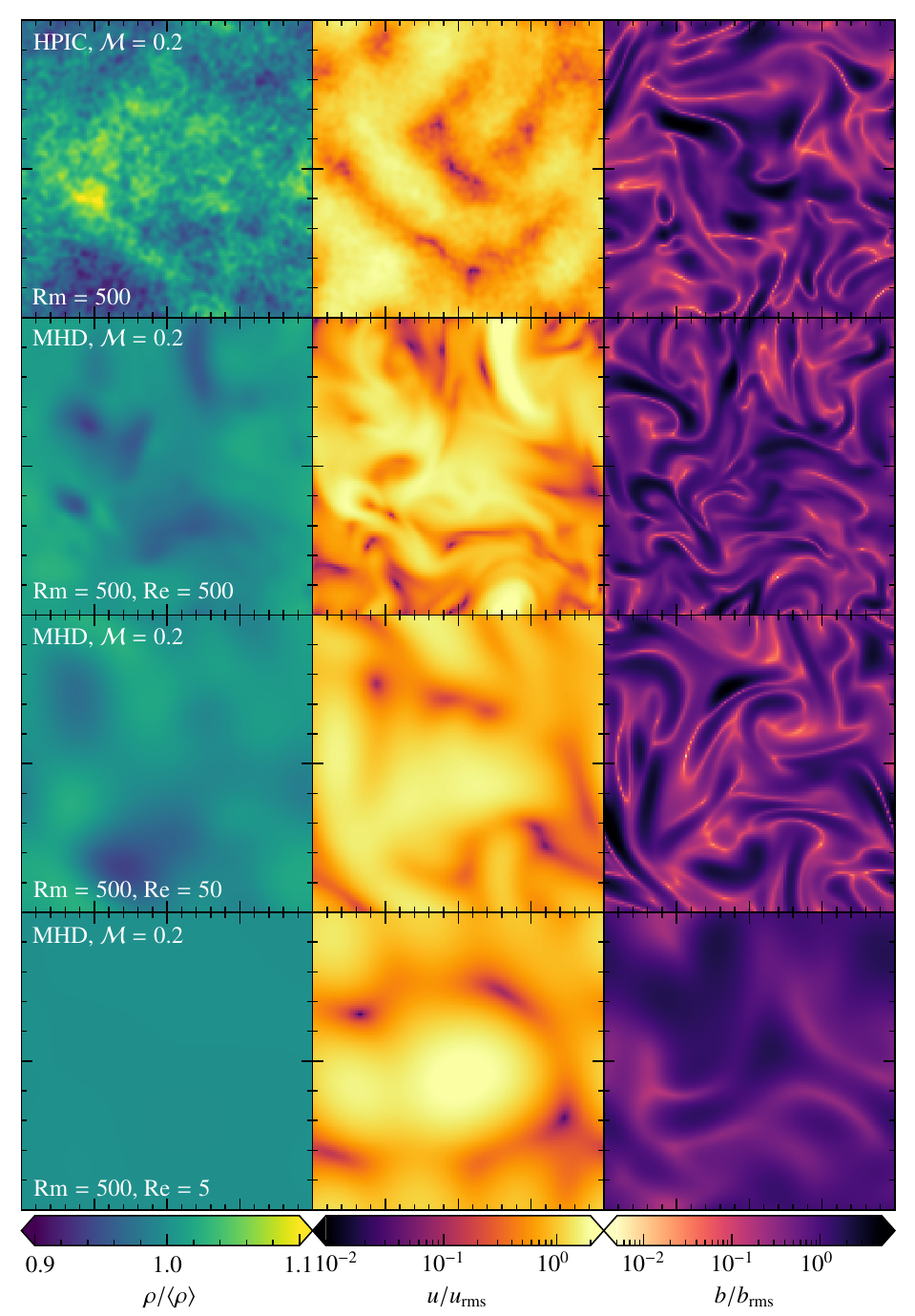}
    \caption{Slice through the mass density, normalised to the mean mass density, $\rho/\langle \rho \rangle$ (first column), the magnitude of velocity, normalised to the root mean square velocity, $u/u_{\rm rms}$ (second column), and the magnitude of the magnetic field, normalised to the root mean square magnetic field, $b/b_{\rm rms}$ (third column), for the subsonic ($\Mach = 0.2$) dynamo simulations in the middle of the kinematic regime ($t = 10\,\ted$). The first row shows HPIC, while the 2nd to 4th row show the MHD runs with $\Re =500$, $50$, and $5$, respectively. We see that the density fluctuations (first column) are small ($\lesssim 10 \%$), as expected for subsonic turbulence. The velocity structures (2nd column) for HPIC are visually similar to the MHD simulation with $\Re=50-500$. The magnetic field structure (3rd column) of HPIC resembles that of the MHD runs with $\Re = 50$ and $\Re = 500$. In the MHD runs, the fluctuations in $\rho$, $u$, and $B$ decrease with decreasing $\Re$, as expected due to the smoothing effect of enhanced viscosity. Apart from the expected small-scale particle noise in the HPIC run, the overall closest visual resemblance is between HPIC and the MHD runs with $\Re\sim 50-500$.}
    \label{fig:slices_subsonic}
\end{figure*}

\Fig{fig:slices_subsonic} shows slice plots of density, velocity, and magnetic field for our HPIC and MHD simulations in the subsonic regime with $\mathcal{M} = 0.2$ in the middle of the kinematic regime ($t = 10\,\ted$). The magnetic energy grows exponentially in the kinematic regime of the turbulent dynamo ($t = 3-17\,\ted$), which is discussed further in \Sec{sec:time_evol}.

The first column shows that the density fluctuations are overall small with $\lesssim 10 \%$ as expected for subsonic plasmas. The HPIC run shows some indication of particle noise ($\lesssim 10 \%$), as expected \citep{AchikanathChirakkara+2024}. For the MHD simulations, we see that the density fluctuations decrease with decreasing $\Re$ due to the increased smoothing of velocity fluctuations with increasing viscosity.

The second column of Fig.~\ref{fig:slices_subsonic} shows that large-scale velocity structures are present in all runs, due to the driving modes being primarily on large scales (c.f., Sec.~\ref{sec:turb_driving}). Similar to the smoothing of density fluctuations, we see that smaller-scale velocity fluctuations decrease as $\Re$ decreases in the MHD simulations. For $\Re = 5$, we find that the velocity structures primarily exist on the driving scale. The velocity structure of HPIC resembles that of the $\Re = 50 - 500$ MHD runs.

The third column of \Fig{fig:slices_subsonic} shows the magnetic field structure. We find small-scale magnetic field fluctuations in HPIC, visually similar to the MHD runs with $\Re = 500$ and $\Re = 50$. For the $\Re = 5$ MHD run, small-scale structures in the magnetic field are absent, and only large-scale fluctuations near the driving scale remain. Comparing HPIC and MHD in the subsonic regime, the closest visual resemblance is between HPIC and the $\Re \sim 50-500$ MHD runs.

\begin{figure*}
    \centering
    \includegraphics[scale=0.88]{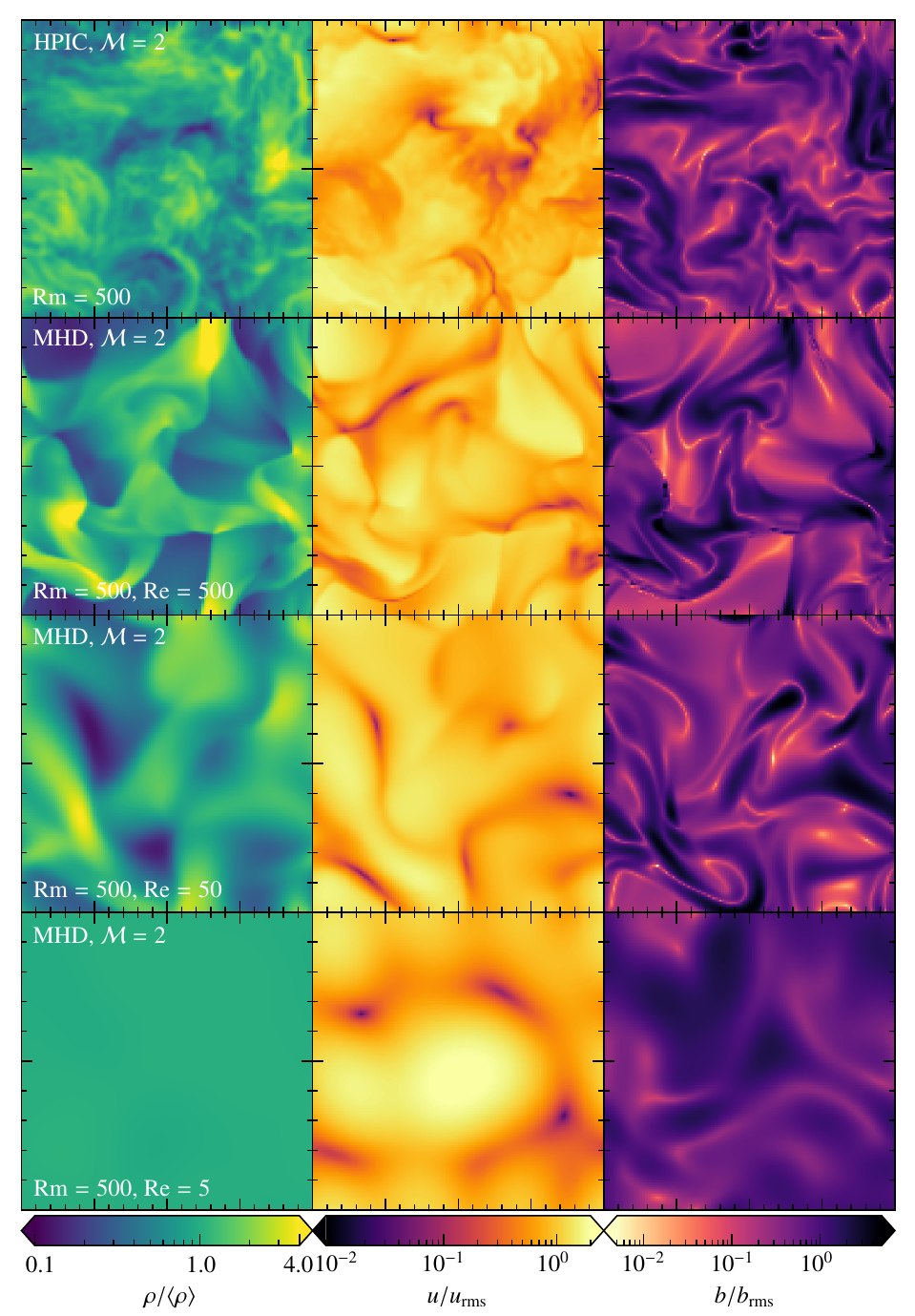}
    \caption{Same as \Fig{fig:slices_subsonic}, but for the supersonic regime ($\Mach = 2$). As expected, we see shock structures and large density contrasts emerging in this regime. Similar to the subsonic runs, also here the turbulent intermediate- and small-scale fluctuations decrease as $\Re$ decreases. The large-scale velocity structures show a significant correlation with the density structures in both HPIC and MHD, while the magnetic field has some large-scale correlation but also significant small-scale structure. The HPIC run is visually closest to the MHD $\Re \sim 50 - 500$ runs in this supersonic regime.}
    \label{fig:slices_supersonic}
\end{figure*}

\Fig{fig:slices_supersonic} shows the corresponding slice plots in the supersonic regime ($\Mach = 2$). The biggest difference compared to the subsonic regime are the strong density fluctuations and the emergence of shocks and large-scale filamentary structure here, as well as the fact that the density and velocity structures have significant correlations, which has been found in previous works on supersonic turbulence \citep{Padoan+1997,Kowal+2007,Kritsuk+2007,Federrath+2010}.
Visually, the supersonic HPIC run is also structurally most similar to the MHD run with $\Re \sim 50 - 500$.

\subsection{Time evolution} \label{sec:time_evol}
\begin{figure*}
\begin{center}
\def\arraystretch{0}
\setlength{\tabcolsep}{0pt}
\begin{tabular}{ll}
\includegraphics[height=0.62\linewidth]{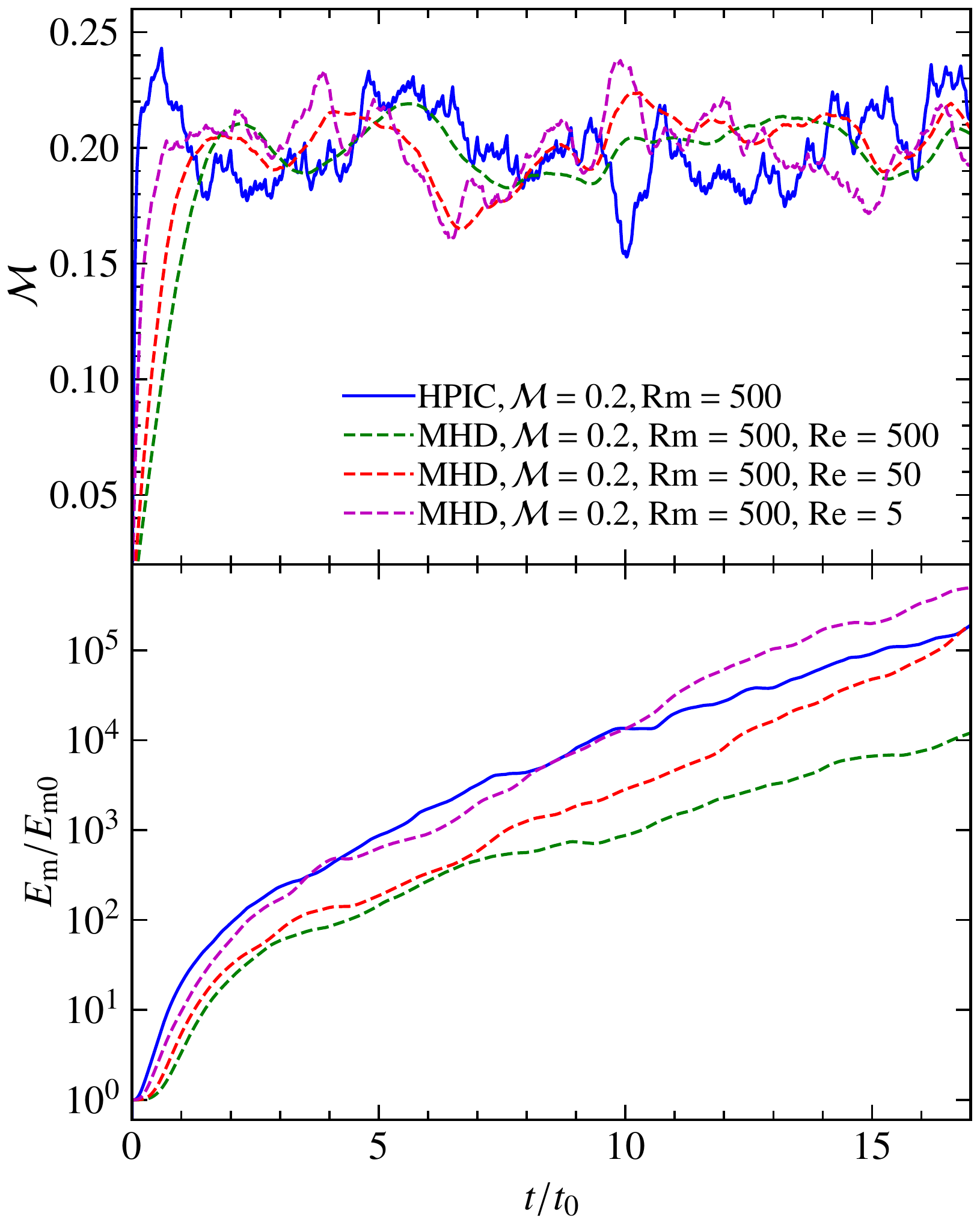} &
\includegraphics[height=0.62\linewidth]{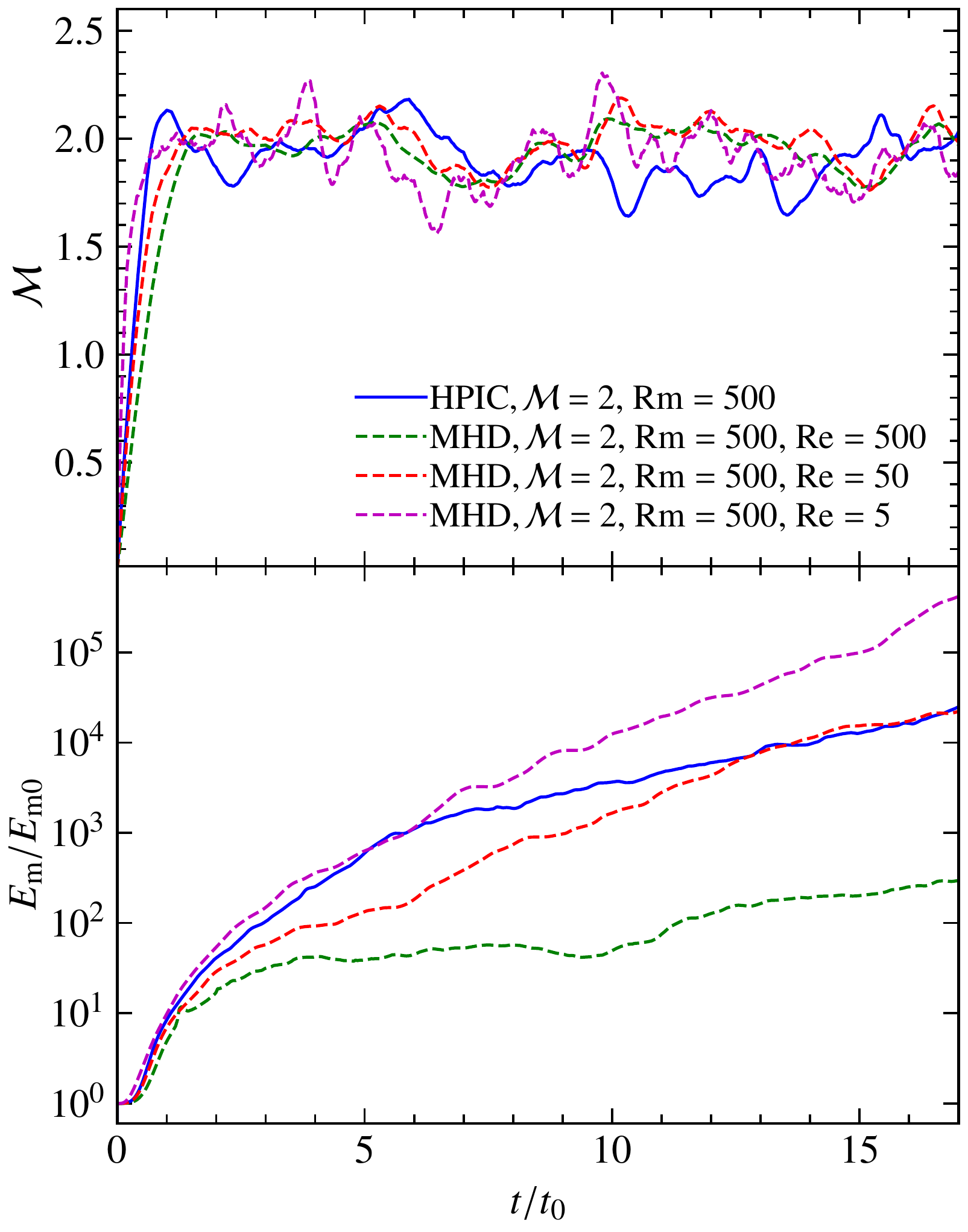} 
\\
\end{tabular}
\end{center}
\caption{The Mach number, $\Mach$, and magnetic energy, normalised to the initial magnetic energy, $\Emag$, as a function of time, normalised to the eddy turn-over time, $\ted$, for the HPIC (solid lines) and MHD turbulent dynamo runs (dashed lines). The left and right panels show the $\Mach = 0.2$ and 2 simulations, respectively. The measured Mach numbers and dynamo growth rates are reported in \Tab{table:sims}. The growth rate of the MHD runs increases as the $\Pm$ of the plasma increases ($\Re$ decreases), for both the $\Mach = 0.2$ and $\Mach = 2$ runs. The growth rate of the HPIC simulation is similar to that of the $\Re = 5 - 500$ MHD runs ($\Pm = 1 - 100$) in the subsonic regime, while in the supersonic regime, the HPIC run's growth rate is similar to that of the $\Re = 50 - 500$ MHD runs ($\Pm = 1 - 10$).}
\label{fig:time_evol}
\end{figure*}

Here we discuss the time evolution of the Mach number ($\Mach$) and the magnetic energy ($\Emag$). The left and right panels of \Fig{fig:time_evol} show the subsonic ($\Mach = 0.2$) and supersonic ($\Mach = 2$) runs, respectively, with the top panels showing $\Mach$ and the bottom panels showing $\Emag$. All runs reach the target steady-state Mach number after a short initial transient phase ($t/\ted\sim0-3$).

The magnetic energy (bottom panels) shows an exponential growth, characteristic of the turbulent dynamo in the kinematic phase. We fit an exponential function, $E_{\rm m} \propto e^{\Gamma t}$, in $t/\ted=3-17$ to determine the growth rate $\Gamma$ in each model. The uncertainties in $\Gamma$ are calculated as described in Appendix~C of \cite{AchikanathChirakkara+2023}, with all fitted values reported in \Tab{table:sims}. As the magnetic Prandtl number of the MHD runs increases, $\Gamma$ increases in both the subsonic and supersonic regimes. The growth rate of the HPIC simulation is similar to that of the $\Re = 5 - 500$ MHD runs ($\Pm = 1 - 100$) in the subsonic regime, whereas in the supersonic regime, it resembles that of the $\Re = 50 - 500$ MHD runs ($\Pm = 1 - 10$). Numerical resolution tests for the growth rate are provided in \App{app:hybridPIC_conv} and \App{app:MHD_conv}, confirming that the $\Gamma$ values reported here are converged.

\subsection{Probability density function} \label{sec:pdfs}
In this section, we study the probability density functions (PDFs) of the density, velocity, and magnetic field.

\subsubsection{Density PDF} \label{sec:dens_pdfs}
\begin{figure*}
\begin{center}
\def\arraystretch{0}
\setlength{\tabcolsep}{0pt}
\begin{tabular}{ll}
\includegraphics[height=0.365\linewidth]{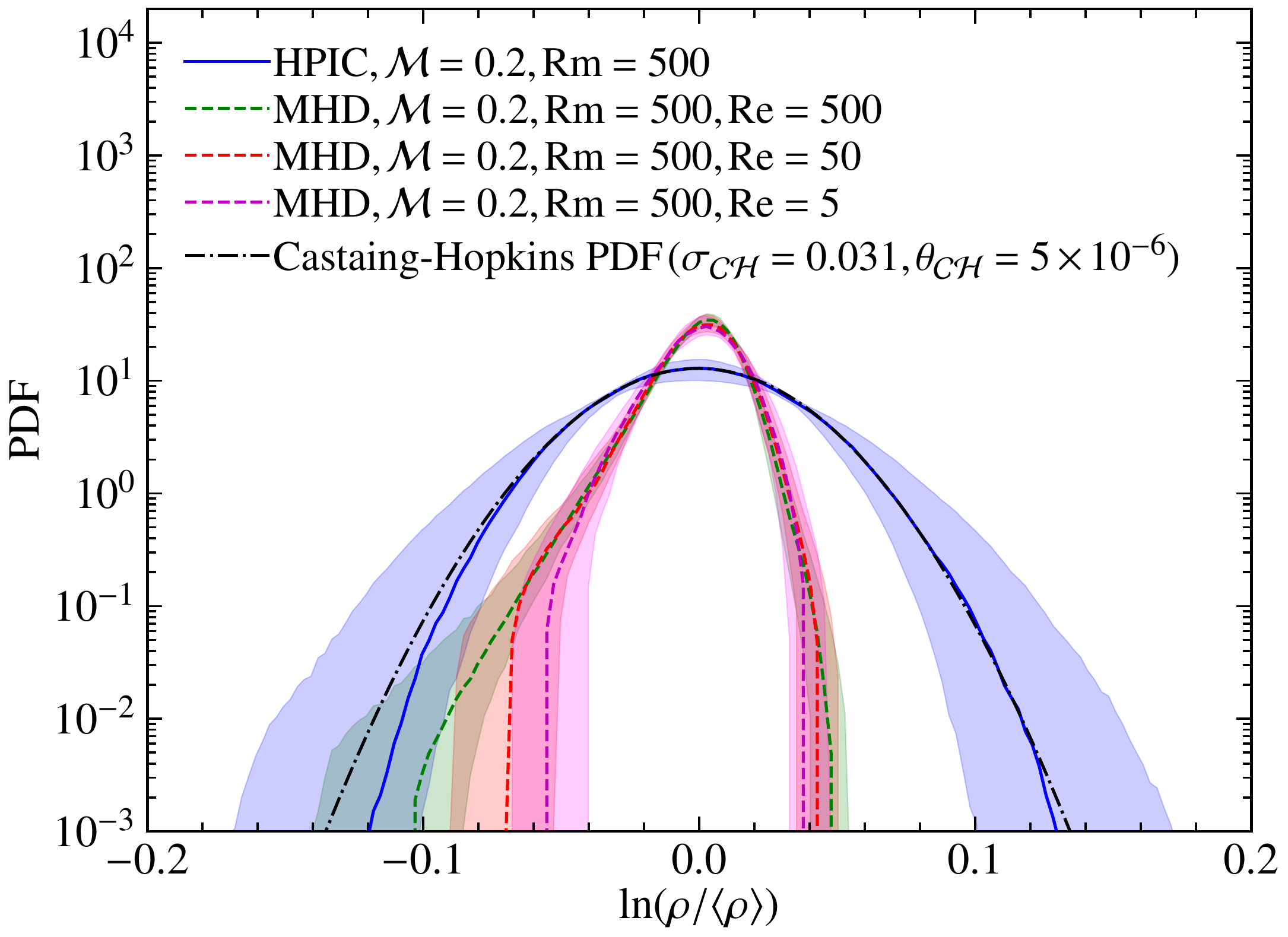} &
\includegraphics[height=0.365\linewidth]{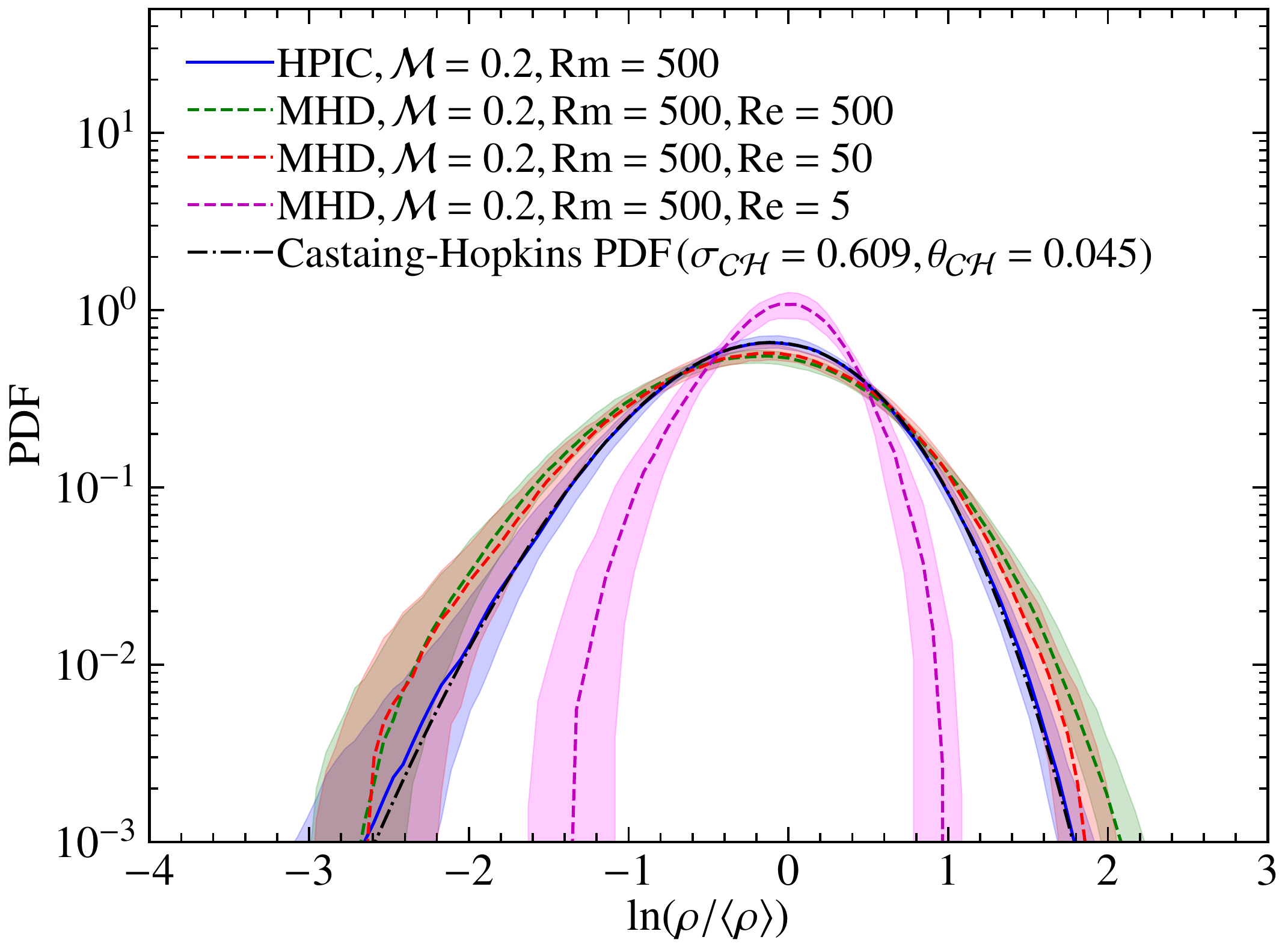}
\end{tabular}
\end{center}
\caption{Probability density functions (PDFs) of the log-density contrast (Eq.~\ref{eq:s}), for HPIC (solid lines) and MHD (dashed lines) with $\Mach = 0.2$ (left panel) and $2$ (right panel), respectively. The PDFs are time-averaged in the kinematic phase of the dynamo ($t/\ted = 3-17$), with the lines and bands representing the median and the $16^{\rm th}$ to $84^{\rm th}$ percentile range, respectively. The magnitude of density fluctuations in the subsonic regime is much smaller than in the supersonic regime. For the MHD runs, we see that as $\Re$ decreases, the density fluctuations decrease due to viscosity suppressing fluctuations in the flow. We fit the density PDFs using the log-normal (\Eq{eqn:lognormal}) and the Castaing-Hopkins (\Eq{eqn:Castaing-Hopkins}) distribution and report the fit parameters in \Tab{table:fits}. The dash-dotted lines show the Castaing-Hopkins model fit for the HPIC simulations, as a representative example, providing a good fit for the supersonic case.}
\label{fig:dens_pdfs}
\end{figure*}

First, we study the density PDFs of the HPIC and MHD simulations in the subsonic and supersonic regimes. \Fig{fig:dens_pdfs} shows the PDFs of the natural logarithm of mass density, $\rho$, normalised to the mean mass density, $\langle \rho \rangle$, time-averaged in the kinematic phase of the dynamo (within time range, $3 \le t/t_0 \le 17$). The data points in all the PDFs shown henceforth represent the median of the time-averaged data, and the lower and upper error bars show the $16^{\rm th}$ and $84^{\rm th}$ percentile, respectively. The $\Mach = 0.2$ and $2$ simulations are shown by the left and right panels, respectively. 

All density PDFs are consistent with Gaussian functions in $\ln(\rho)$ (log-normal in $\rho$), as expected for turbulent flows \citep{VazquezSemadeni1994,Padoan+1997, Passot&VazquezSemadeni1998,Kritsuk+2007,Federrath+2008,Konstandin+2012}. The more viscous MHD runs (i.e., the runs with lower $\Re$) show narrower distributions, which reflects the suppression of density fluctuations in highly viscous plasmas. The density fluctuations are substantially larger (by orders of magnitude) in $\Mach=2$ compared to $\Mach=0.2$ due to the development of shocks in the supersonic regime. The density fluctuations in the supersonic HPIC run agree with those in the respective $\Re=50-500$ MHD runs. In contrast, the subsonic HPIC run has a somewhat larger dispersion compared to the respective MHD runs, due to PIC noise dominating the density fluctuations in the subsonic regime, as we saw in the density slices (c.f.~top left panel of Fig.~\ref{fig:slices_subsonic}).

To quantify the shape of the density PDFs, we fit a log-normal and a Castaing-Hopkins model \citep{Castaing1996, Hopkins2013, Federrath&Banerjee2015, Seta+2023} to the PDFs, in the log-density contrast \citep{Passot&VazquezSemadeni1998, Kritsuk+2007,Federrath+2010},
\begin{equation} \label{eq:s}
q = \ln(\rho / \langle \rho \rangle).
\end{equation}

\noindent The log-normal ($\mathcal{LN}$) PDF is defined as
\begin{equation}\label{eqn:lognormal}
    \mathcal{LN}(q) = \left(2\pi\sigma_{\mathcal{LN}}^{2}\right)^{-1/2} \exp\left[-\frac{1}{2} \left(\frac{q - \mu}{\sigma_{\mathcal{LN}}}\right)^{2}\right],
\end{equation}
where $\mu = -\sigma^2_\mathcal{LN}/2$ is the mean and $\sigma^2_{\mathcal{LN}}$ is the variance. The Castaing-Hopkins ($\mathcal{CH}$) PDF is defined as
\begin{equation} \label{eqn:Castaing-Hopkins}
    \mathcal{CH}(q) = I_{1} \left(2 (\lambda \omega (q))^{-1/2}\right) \exp \left[ -(\lambda + \omega(q))  \right] \left( \frac{\lambda}{\theta_{\mathcal{CH}}^{2} \omega(q)} \right)^{-1},
\end{equation}
where $\lambda = \sigma_{\mathcal{CH}}^{2}/(2\theta_{\mathcal{CH}}^{2})$ and $\omega(q) = \lambda/(1 + \theta_{\mathcal{CH}}) - q/\theta_{\mathcal{CH}}$. Here $I_{1}$ is the modified Bessel function of the first kind, $\sigma^2_{\mathcal{CH}}$ is the variance, and $\theta_{\mathcal{CH}}$ is the intermittency parameter. The larger $\theta_{\mathcal{CH}}$, the stronger the intermittency. For $\theta_{\mathcal{CH}} \rightarrow 0$, the Castaing-Hopkins model reduces to the log-normal distribution.

All fit parameters are listed in \Tab{table:fits}. We find that the intermittency parameter, $\theta_{\mathcal{CH}} \rightarrow 0$, for the $\Mach=0.2$ runs, and hence $\sigma_{\mathcal{CH}} \sim \sigma_{\mathcal{LN}}$, confirming that the log-normal is a good model in the subsonic regime. The intermittency parameter is much larger in the supersonic regime, as found in previous MHD simulations \citep{Hopkins2013,Federrath&Banerjee2015}. However, at $\Mach=2$, $\theta_{\mathcal{CH}}$ is still relatively small, indicating that the log-normal is a reasonable model for density fluctuations in the mildly supersonic regime.

\begin{table*}
\centering
\caption{Fit parameters for the density (${\rm ln}(\rho/\langle \rho \rangle)$) PDFs, and the $x$-components of the velocity ($u_x/u_{\rm rms}$) and magnetic field ($b_x/b_{\rm rms}$) PDFs.}
\begin{tabular}{lc|ccc|cc|ccc}
\toprule
& \multicolumn{3}{|c|}{${\rm ln}(\rho/\langle \rho \rangle)$} & \multicolumn{2}{c}{$u_{x}/u_{\rm rms}$} & \multicolumn{3}{c}{$b_{x}/b_{\rm rms}$} \\
\cmidrule(lr){2-4}
\cmidrule(lr){5-6}
\cmidrule(lr){7-9}
Model & $\sigma_{\mathcal{LN}}$ & $\sigma_{\mathcal{CH}}$ & $\theta_{\mathcal{CH}}$ & $\mu_{\mathcal{N}}$ & $\sigma_{\mathcal{N}}$ & $\mu_{\mathcal{CN}}$ & $\sigma_{\mathcal{CN}}$ & $\gamma_{\mathcal{CN}}$ \\
\bottomrule
\texttt{HPICM0.2Rm500} & $0.031_{\minus0.001}^{\plus0.001}$ & $0.031_{\minus0.001}^{\plus0.001}$ & ${5\times10^{-6}}_{\minus4 \times 10^{-6}}^{\plus6 \times 10^{-4}}$ & $-0.002_{\minus0.012}^{\plus0.013}$ & $0.573_{\minus0.012}^{\plus0.012}$ & $0.000_{\minus0.001}^{\plus0.007}$ & $2.670_{\minus0.170}^{\plus1.817}$ & $0.224_{\minus0.022}^{\plus0.076}$ \\
\\
\texttt{MHDM0.2Rm500Re500} & $0.012_{\minus0.001}^{\plus0.001}$ & $0.012_{\minus0.001}^{\plus0.001}$ & $0.002_{\minus0.001}^{\plus0.001}$ & $\phantom{-}0.005_{\minus0.009}^{\plus0.008}$ & $0.591_{\minus0.007}^{\plus0.007}$ & $0.001_{\minus0.001}^{\plus0.004}$ & $1.382_{\minus0.153}^{\plus0.261}$ & $0.351_{\minus0.018}^{\plus0.014}$ \\
\\
\texttt{MHDM0.2Rm500Re50} & $0.013_{\minus0.001}^{\plus0.001}$ & $0.013_{\minus0.001}^{\plus0.001}$ & $0.002_{\minus0.001}^{\plus0.001}$ & $-0.017_{\minus0.019}^{\plus0.017}$ & $0.596_{\minus0.015}^{\plus0.016}$ & $0.003_{\minus0.003}^{\plus0.009}$ & $2.142_{\minus0.499}^{\plus1.501}$ & $0.272_{\minus0.021}^{\plus0.021}$ \\
\\
\texttt{MHDM0.2Rm500Re5} & $0.014_{\minus0.001}^{\plus0.001}$ & $0.014_{\minus0.001}^{\plus0.001}$ & $0.002_{\minus0.001}^{\plus0.001}$ & $\phantom{-}0.000_{\minus0.015}^{\plus0.016}$ & $0.580_{\minus0.013}^{\plus0.015}$ & $0.001_{\minus0.001}^{\plus0.005}$ & $2.548_{\minus0.732}^{\plus1.676}$ & $0.237_{\minus0.016}^{\plus0.020}$ \\
\\
\texttt{HPICM2Rm500} & $0.595_{\minus0.009}^{\plus0.009}$ & $0.609_{\minus0.011}^{\plus0.011}$ & $0.045_{\minus0.010}^{\plus0.010}$ &  $\phantom{-}0.000_{\minus0.015}^{\plus0.013}$ & $0.580_{\minus0.013}^{\plus0.012}$ & $0.000_{\minus0.001}^{\plus0.005}$ & $2.085_{\minus0.085}^{\plus1.501}$ & $0.242_{\minus0.027}^{\plus0.058}$ \\
\\
\texttt{MHDM2Rm500Re500} & $0.709_{\minus0.010}^{\plus0.011}$ & $0.725_{\minus0.012}^{\plus0.012}$ & $0.045_{\minus0.012}^{\plus0.013}$ &  $-0.027_{\minus0.011}^{\plus0.012}$ & $0.588_{\minus0.012}^{\plus0.011}$ & $0.001_{\minus0.008}^{\plus0.007}$ & $2.543_{\minus0.615}^{\plus1.239}$ & $0.242_{\minus0.016}^{\plus0.015}$ \\
\\
\texttt{MHDM2Rm500Re50} & $0.683_{\minus0.013}^{\plus0.013}$ & $0.703_{\minus0.016}^{\plus0.014}$ & $0.056_{\minus0.015}^{\plus0.018}$ & $-0.016_{\minus0.014}^{\plus0.015}$ & $0.590_{\minus0.015}^{\plus0.014}$ & $0.000_{\minus0.001}^{\plus0.008}$ & $2.529_{\minus0.529}^{\plus2.462}$ & $0.237_{\minus0.024}^{\plus0.063}$ \\
\\
\texttt{MHDM2Rm500Re5} & $0.367_{\minus0.016}^{\plus0.018}$ & $0.376_{\minus0.006}^{\plus0.007}$ & $0.043_{\minus0.006}^{\plus0.007}$ & $\phantom{-}0.017_{\minus0.015}^{\plus0.016}$ & $0.576_{\minus0.015}^{\plus0.017}$ & $0.005_{\minus0.005}^{\plus0.007}$ & $2.105_{\minus0.217}^{\plus0.907}$ & $0.246_{\minus0.019}^{\plus0.054}$ \\
 \hline
\end{tabular}
\begin{tablenotes}
    \item Notes: The first three columns depict the lognormal (\Eq{eqn:lognormal}) and Castaing-Hopkins (\Eq{eqn:Castaing-Hopkins}) fit parameters for the density PDFs. The velocity PDFs are fit using a normal distribution and the fit parameters are shown in the fourth and fifth columns. The last three columns show the Cauchy-normal (\Eq{eqn:cauchy-normal}) fit parameters for the magnetic field PDFs.
\end{tablenotes}
\label{table:fits}
\end{table*}

\subsubsection{Velocity PDF} \label{sec:vels_pdfs}

\begin{figure*}
\begin{center}
\def\arraystretch{0}
\setlength{\tabcolsep}{0pt}
\begin{tabular}{ll}
\includegraphics[height=0.365\linewidth]{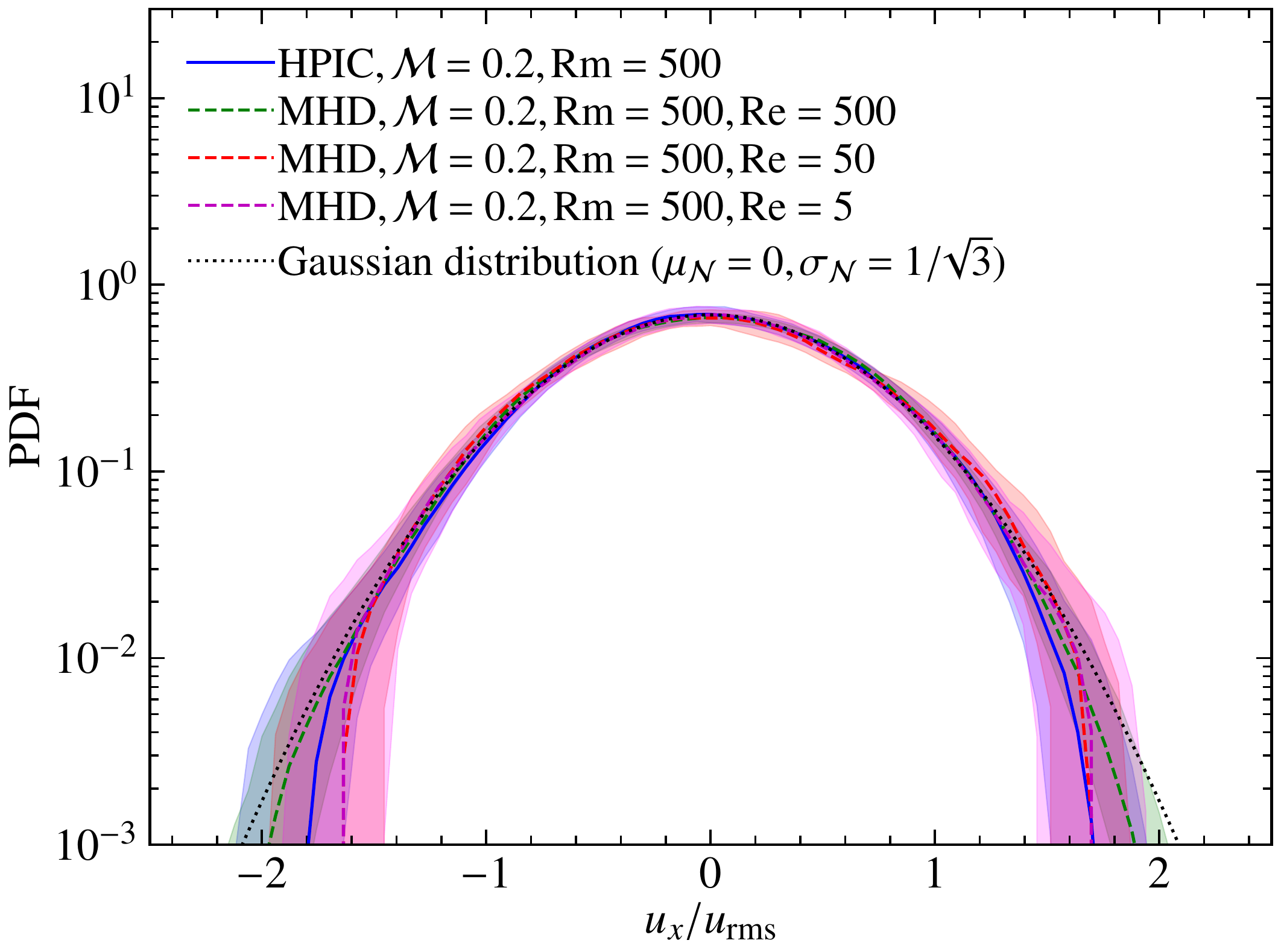} &
\includegraphics[height=0.365\linewidth]{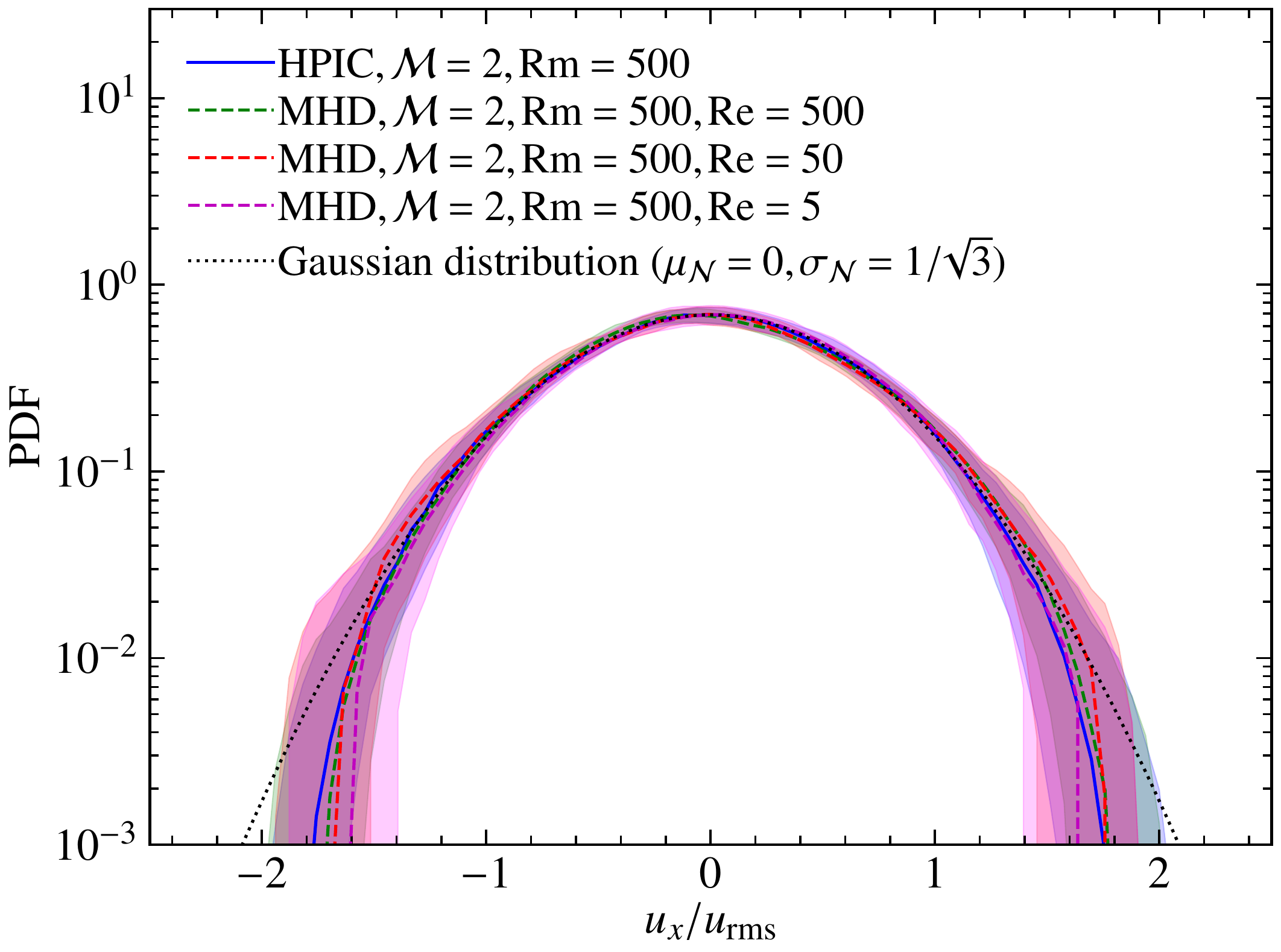}
\end{tabular}
\end{center}
\caption{Same as \Fig{fig:dens_pdfs}, but for the $x$-component of the velocity, $u_{x}$, normalised to the root mean square velocity, $u_{\rm rms}$. The dotted line shows a Gaussian function with mean, $\mu_{\mathcal{N}} = 0$, and standard deviation, $\sigma_{\mathcal{N}} = 1/\sqrt{3}$, for comparison. The velocity PDFs of all the simulations show good agreement with this Gaussian function.}
\label{fig:vels_pdfs}
\end{figure*}

\Fig{fig:vels_pdfs} shows the same as \Fig{fig:dens_pdfs}, but for the $x$-component of the velocity, $u_x$, normalised to the root mean square velocity, $u_{\rm rms}$. We fit the velocity PDFs with a Gaussian (normal; $\mathcal{N}$) distribution by using $q = u_x/u_{\rm rms}$ in \Eq{eqn:lognormal}, with $\mu_{\mathcal{N}}$ and $\sigma_{\mathcal{N}}$ as the mean and standard deviation, respectively. The fit parameters are listed in \Tab{table:fits}. The velocity PDFs of all simulations are in good agreement with a Gaussian distribution with $\mu_{\mathcal{N}} \sim 0$ and $\sigma_{\mathcal{N}} \sim 1/\sqrt{3}$, as expected for isotropic turbulence \citep[e.g., see appendix in][]{Federrath+2013}. While we only show the $x$-component here, all 3~velocity components have very similar PDFs. This is indeed reflected in the fact that $\sigma_{\mathcal{N}} \sim 1/\sqrt{3}$ of a single component, i.e., when all 3~components are combined, we recover the target Mach number in each respective simulation.
Finally, varying $\Re$ in the MHD runs does not significantly impact the shape of the velocity PDFs. However, increasing the viscosity would reduce $u_{\rm rms}$ for the same turbulence driving amplitude. Therefore, we adjust the driving amplitude (see \Sec{sec:turb_driving}) to ensure that the same target Mach numbers are reached for all choices of viscosity. The driving amplitude is fixed and does not vary dynamically throughout the simulation.

\subsubsection{Magnetic field PDF} \label{sec:mags_pdfs}

\begin{figure*}
\begin{center}
\def\arraystretch{0}
\setlength{\tabcolsep}{0pt}
\begin{tabular}{ll}
\includegraphics[height=0.365\linewidth]{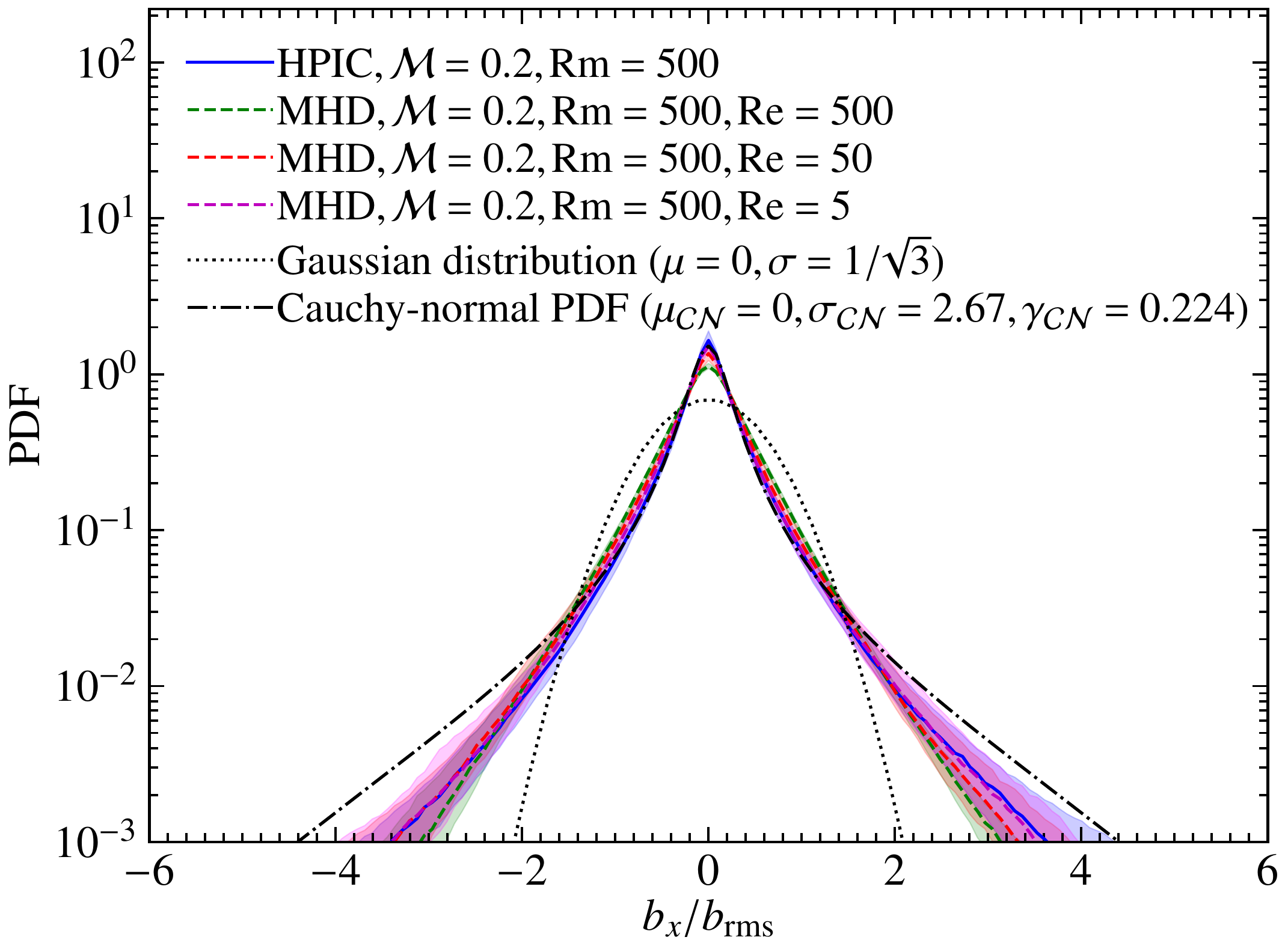} &
\includegraphics[height=0.365\linewidth]{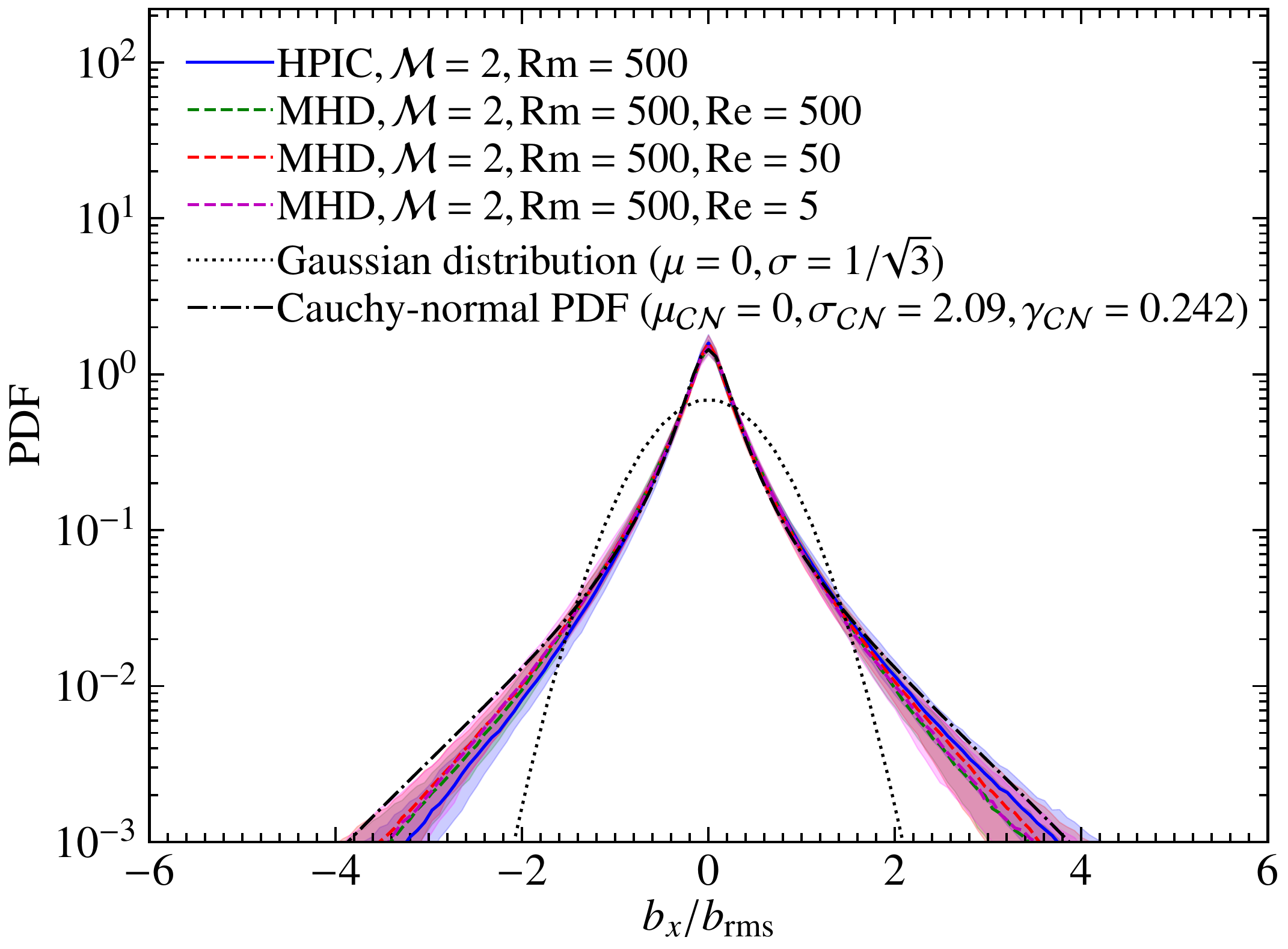}
\end{tabular}
\end{center}
\caption{Same as \Fig{fig:dens_pdfs}, but for the $x$-component of the magnetic field, $b_{x}$, normalised to the root mean square magnetic field, $b_{\rm rms}$. The dash-dotted lines show the Cauchy-normal fit (\Eq{eqn:cauchy-normal}) for the HPIC simulations. The dotted line shows a Gaussian (\Eq{eqn:lognormal}) with $\mu = 0$ and $\sigma = 1/\sqrt{3}$ for comparison. The magnetic fields are clearly non-Gaussian (intermittent) in both the subsonic and supersonic regimes and show reasonable agreement with the Cauchy-normal function.}
\label{fig:mags_pdfs}
\end{figure*}

Finally, we study the magnetic field PDFs of the HPIC and MHD simulations. The PDFs of the $x$-component of the magnetic field, $b_x$, normalised to the root mean square magnetic field value, $b_{\rm rms}$ are shown in \Fig{fig:mags_pdfs}. We see that the magnetic field PDFs for all the simulations are non-Gaussian, with strong peaks around $b=0$, reflecting the fact that mean fields are not present, and with very strong tails towards extreme $b$ values. We model these PDFs using a Cauchy-normal distribution \citep{Seta+2021},
\begin{equation} \label{eqn:cauchy-normal}
    \mathcal{CN} (q) = A {\left[1 + \left(\frac{x - \mu_{\mathcal{CN}}}{\gamma_{\mathcal{CN}}}\right)^{2}\right]}^{-1} \exp\left[-\frac{1}{2} \left(\frac{x - \mu_{\mathcal{CN}}}{\sigma_{\mathcal{CN}}}\right)^{2}\right],
\end{equation}
 where $q=b_{x}/b_{\rm rms}$ and $\mu_{\mathcal{CN}}$, $\sigma_{\mathcal{CN}}$, and $\gamma_{\mathcal{CN}}$ are the mean, the standard deviation of the normal part, and the scale of the Cauchy part, respectively. The coefficient
\begin{equation}
    A = \frac{1}{2\pi^{2}\gamma_{\mathcal{CN}}} \exp\left[-\left( \frac{\gamma_{\mathcal{CN}}}{2\sigma_{\mathcal{CN}}} \right)^{2} \right] \left[ 1 - {\rm erf} \left( \frac{\gamma_{\mathcal{CN}}}{\sqrt{2}\sigma_{\mathcal{CN}}} \right) \right]^{-1},
\end{equation}
where ${\rm erf}$ is the error function, ensuring that the PDF is normalised, i.e., $\int\mathcal{CN}(q)dq=1$. Using this distribution function, we fit the magnetic field PDFs for all runs and list the fit parameters in \Tab{table:fits}. The fits provide good approximations of the simulation data around the peak of the PDFs, with the tails being somewhat overestimated.

\subsection{Spectra} \label{sec:spectra}

In order to understand the characteristics of the plasma on different scales, we study the power spectra of magnetic energy, electric current, and turbulent kinetic energy.

\subsubsection{Magnetic energy spectra} \label{sec:mags_spectra}
\begin{figure*}
\begin{center}
\def\arraystretch{0}
\setlength{\tabcolsep}{0pt}
\begin{tabular}{ll}
\includegraphics[height=0.375\linewidth]{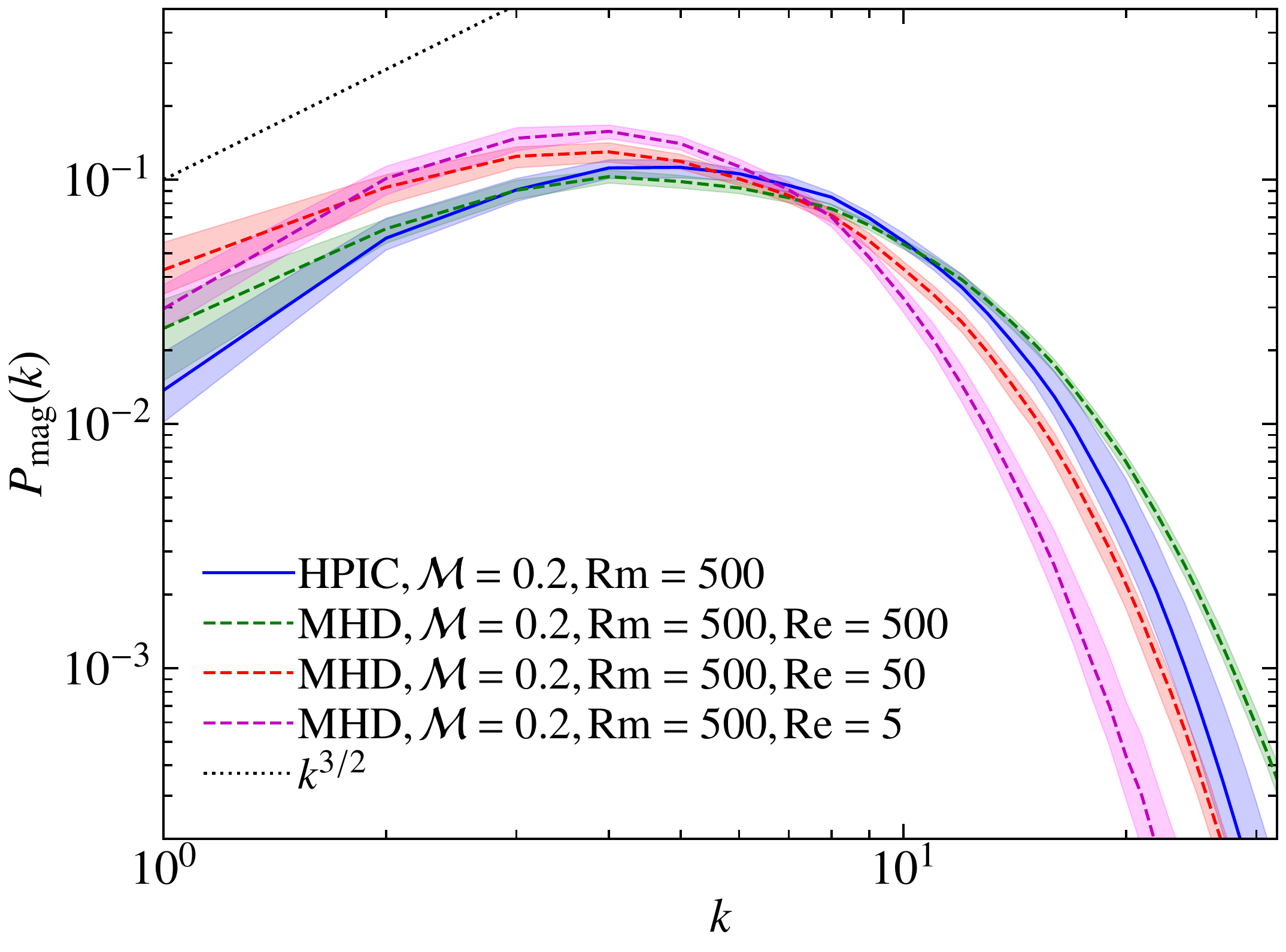} &
\includegraphics[height=0.375\linewidth]{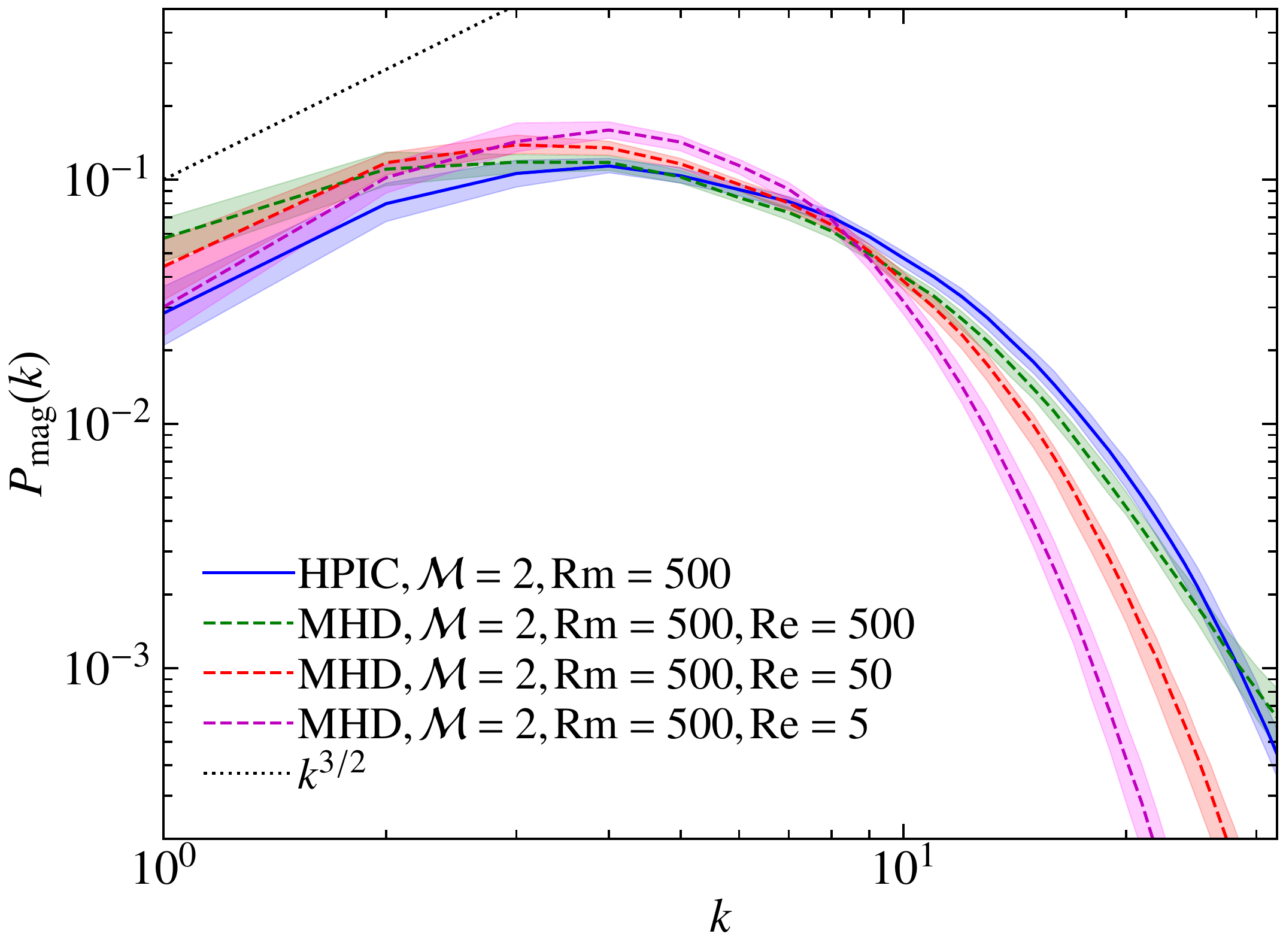} 
\\
\end{tabular}
\end{center}
\caption{Magnetic energy power spectra for the HPIC (solid lines) and MHD (dashed lines) simulations with $\Mach = 0.2$ (left panel) and $\Mach = 2$ (right panel). The spectra presented here are time-averaged in the kinematic phase of the dynamo ($t = 3 - 17 \ted$), and normalised to the total energy. The lines are the median ($50^{\rm th}$ percentile) and the shaded regions range from the $16^{\rm th}$ to the $84^{\rm th}$ percentile, showing the time variation. The dotted line shows a $k^{3/2}$ scaling \citep{Kazantsev1968} for comparison. From the MHD runs, we see that as $\Re$ decreases, the magnetic energy moves to somewhat larger scales (lower $k$). Here HPIC is most similar to the MHD runs with $\Re\sim50-500$.}
\label{fig:mags_spectra}
\end{figure*}

The magnetic energy power spectra of our simulations are shown in \Fig{fig:mags_spectra}, with $\Mach = 0.2$ on the left and $\Mach = 2$ on the right. All the spectra shown henceforth are normalised such that the total power is unity and are also time-averaged in the kinematic regime of the dynamo. The lines represent the $50^{\rm th}$ percentile (median) of the data, and the shaded regions bracket the $16^{\rm th}$ to $84^{\rm th}$ percentile range. The $k^{3/2}$ scaling (dotted line), characteristic of the MHD turbulent dynamo in incompressible flows during the kinematic regime \citep{Kazantsev1968, Seta+2020}, also appears in the kinematic phase of compressive simulations \citep{Seta&Federrath2020, Seta&Federrath2021b}. This scaling shows reasonable agreement with all the simulations, both in the subsonic and supersonic regimes. However, the limited dynamic range at large scales (low $k$) prevents a detailed quantification of the scaling behaviour at those scales. \citet{Federrathetal2014ApJ} also find the $k^{3/2}$ scaling at much higher Mach numbers, $\Mach = 11$, in MHD simulations with a large magnetic Prandtl number, $\Pm = 10$. For the MHD runs, the magnetic energy shifts to larger scales as $\Re$ decreases. The magnetic spectra for HPIC in both the subsonic and supersonic regimes are most similar to those of the respective MHD runs with $\Re \sim 50-500$, although there are some significant differences across all scales.

\subsubsection{Electric current spectra} \label{sec:current_spectra}
\begin{figure*}
\begin{center}
\def\arraystretch{0}
\setlength{\tabcolsep}{0pt}
\begin{tabular}{ll}
\includegraphics[height=0.375\linewidth]{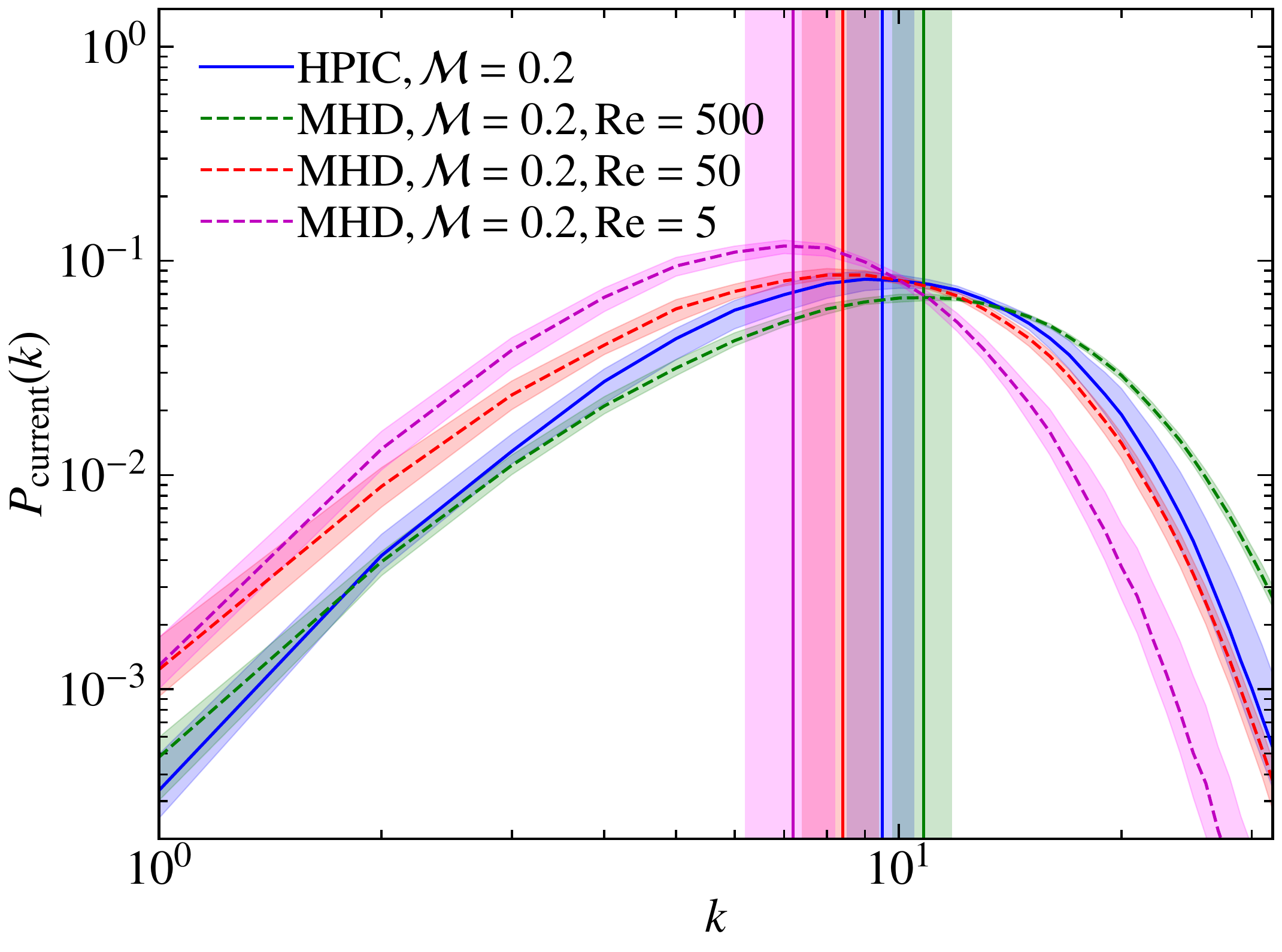} &
\includegraphics[height=0.375\linewidth]{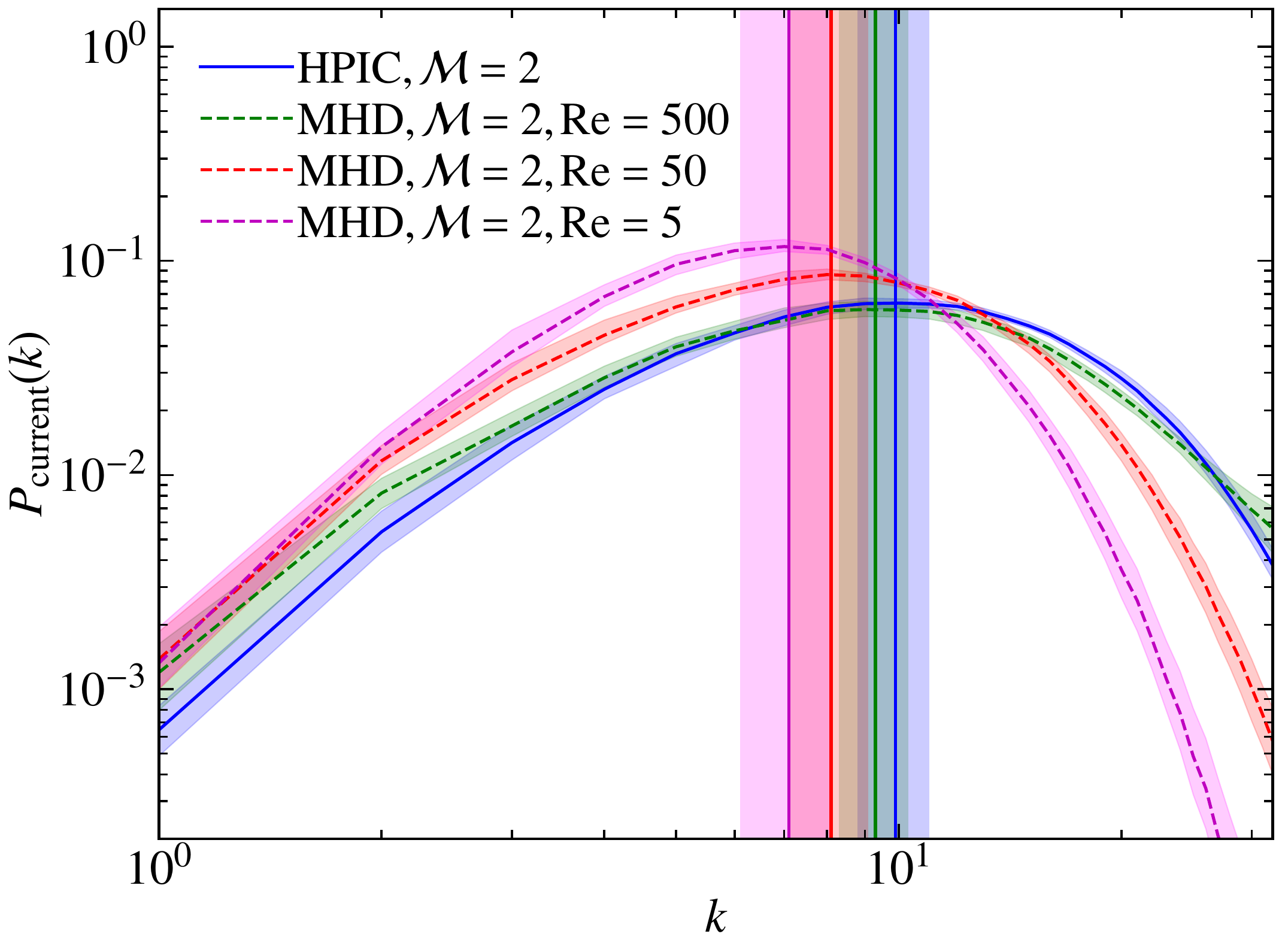} 
\\
\end{tabular}
\end{center}
\caption{Same as \Fig{fig:mags_spectra} but for the total current. The peak of the current spectra corresponds to the scale of magnetic energy dissipation, $k_{\eta}$, indicated by the vertical solid lines, with the shaded region representing the error. The values of $k_{\eta}$ are reported in \Tab{table:sims}. For the MHD runs, $k_{\eta}$ decreases with $\Re$. The $k_{\eta}$ for the $\Mach = 0.2$ HPIC run is comparable to that of the MHD simulation with $\Re = 50$–$500$, while for the $\Mach = 2$ HPIC run, it is similar to the MHD run with $\Re \sim 500$.}
\label{fig:current_spectra}
\end{figure*}

\Fig{fig:current_spectra} shows the spectra of the total current, $\J = (\nabla \times \B)/\mu_{0}$. Following the definition in \citet{Kriel+2023}, the peak wavenumber of the current spectra are used to represent the dissipation scale of the magnetic energy, i.e., the resistive wavenumber $k_{\eta}$. As in \citet{Kriel+2023}, the spectra are smoothed using a third-order cubic spline interpolation operator before obtaining $k_{\eta}$. The reported value and error are the mean and standard deviation of the peak wavenumber, as listed in \Tab{table:sims}. The $k_{\eta}$ for the MHD runs decreases with $\Re$, similar to the magnetic energy, which shifts to larger length scales as $\Re$ decreases. The magnetic dissipation scale for the $\Mach = 0.2$ HPIC run is comparable to that of the MHD simulation with $\Re = 50$–$500$. For the $\Mach = 2$ HPIC run, it is more similar to the MHD run with $\Re \sim 500$.

\subsubsection{Turbulent kinetic energy spectra} \label{sec:vels_spectra}

\begin{figure*}
\begin{center}
\def\arraystretch{0}
\setlength{\tabcolsep}{0pt}
\begin{tabular}{ll}
\includegraphics[height=0.375\linewidth]{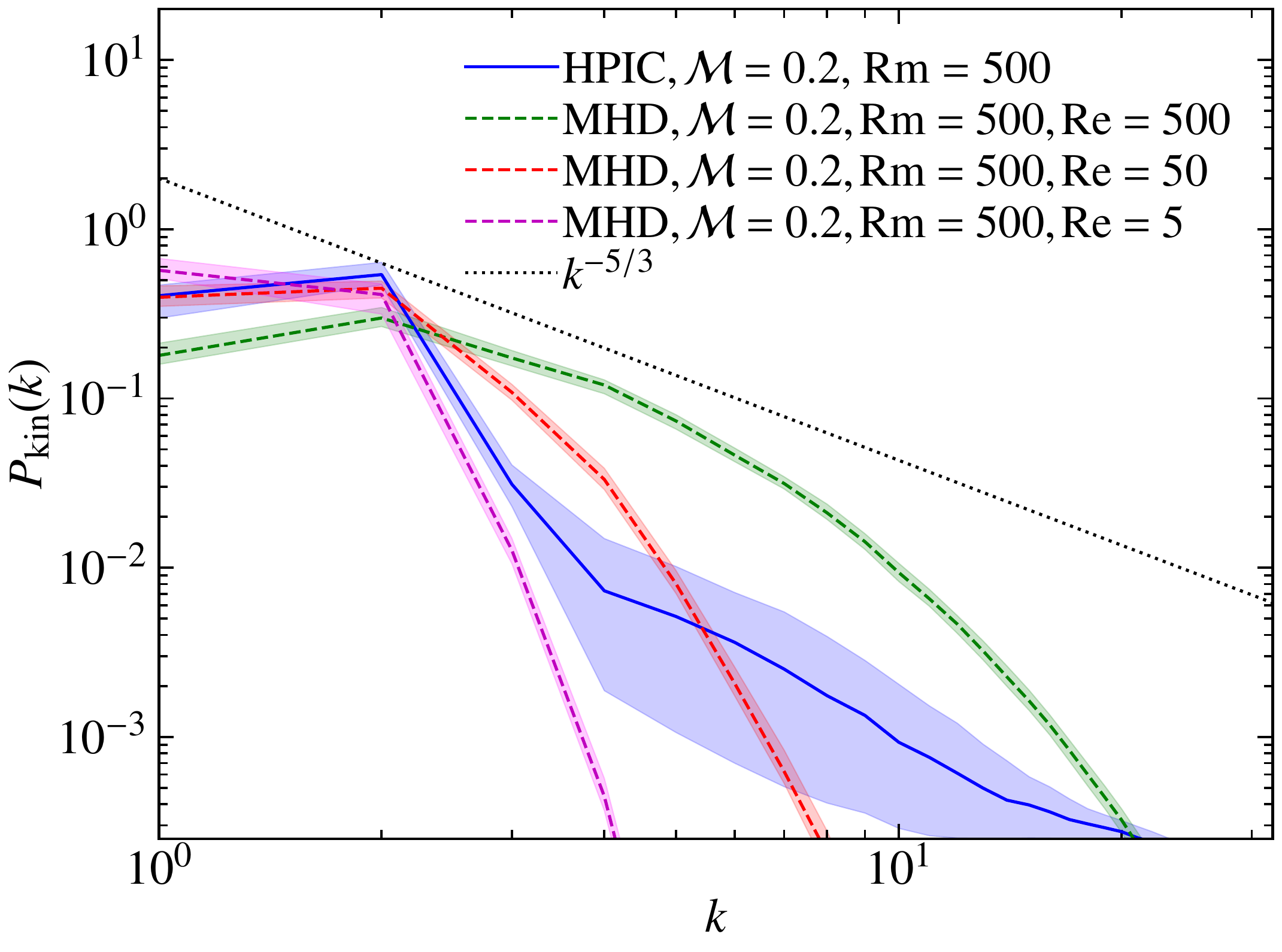} &
\includegraphics[height=0.375\linewidth]{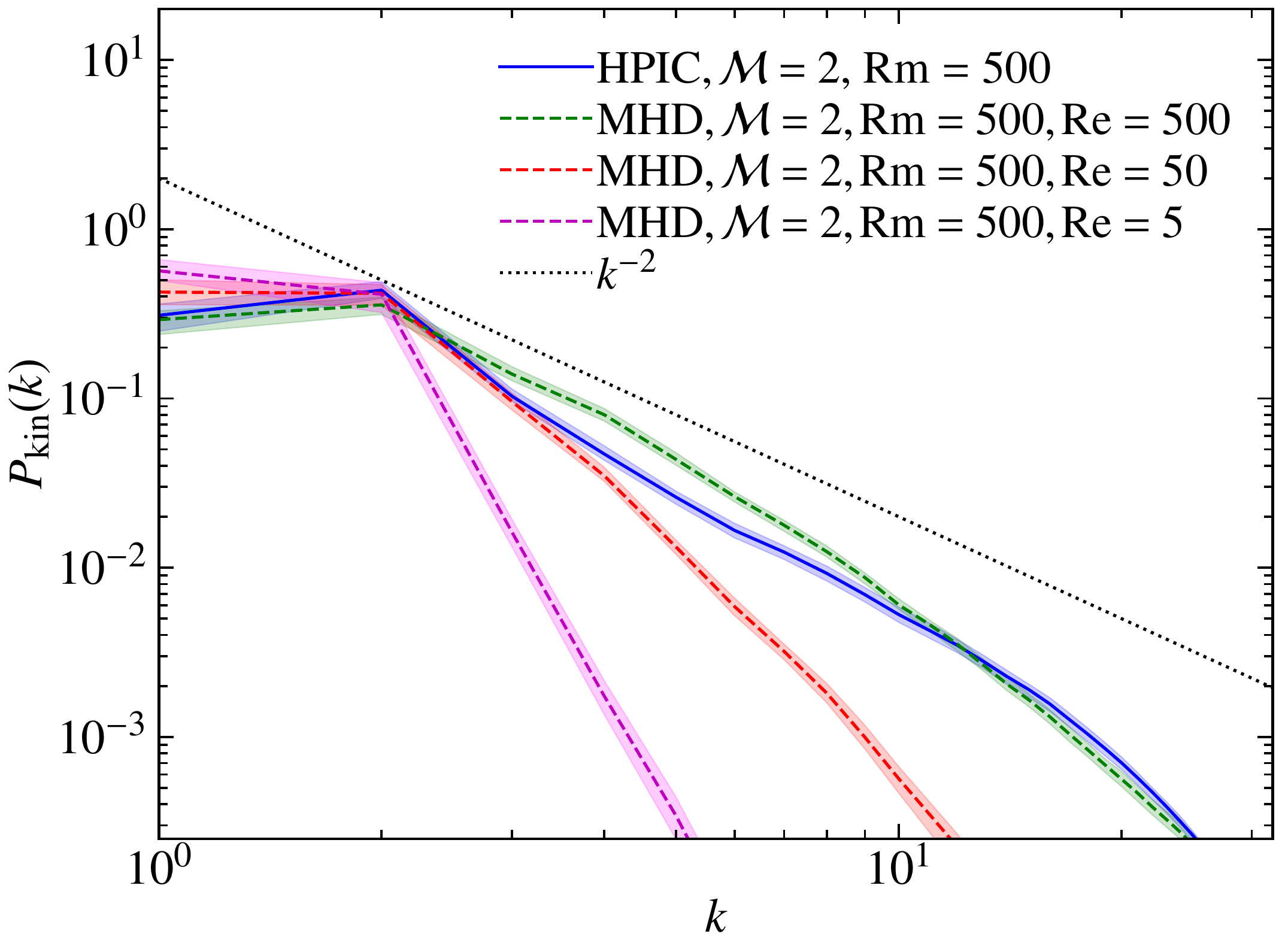} 
\\
\end{tabular}
\end{center}
\caption{Same as \Fig{fig:mags_spectra} but for the turbulent kinetic energy. For comparison, we show the $k^{-5/3}$ Kolmogorov scaling, characteristic of subsonic turbulence (left panel) and the $k^{-2}$ Burgers scaling (dotted lines), characteristic of supersonic turbulence (right panel). As $\Re$ decreases, both the scaling range and the dissipation wavenumber, $k_{\nu}$, decrease. The two MHD simulations with $\Re = 500$ agree reasonably with the Kolmogorov scaling in the subsonic regime, and the Burgers scaling in the supersonic regime, respectively. The spectrum of the subsonic HPIC run resembles that of the MHD run with $\Re \sim 50-500$. In the supersonic regime, the HPIC run resembles the MHD run with $\Re = 500$.}
\label{fig:vels_spectra}
\end{figure*}

The power spectra of the turbulent kinetic energy are shown in \Fig{fig:vels_spectra}. The $k^{-5/3}$ Kolmogorov scaling, characteristic of subsonic turbulence and the $k^{-2}$ Burgers scaling relation, characteristic of supersonic turbulence, are shown as guidelines for the $\Re=500$ MHD runs. The maximum turbulent energy is injected at $k_{\rm turb} = 2$, as confirmed by the spectra. As the $\Re$ of the MHD simulations is decreased in both the $\Mach=0.2$ and $\Mach=2$ runs, the kinetic energy spectra get steeper, and the scaling range becomes shorter -- it is practically absent in the $\Re = 50$ and $5$ runs, as expected, due to the low Reynolds numbers of these flows not allowing turbulent cascade to develop. The dissipation scale of the turbulence shifts to larger length scales (smaller $k$) as the viscosity of the plasma increases. In the subsonic regime, the $\Pkin$ spectrum of HPIC resembles that of the $\Re \sim 50 - 500$ MHD run. The spectrum of the supersonic HPIC run closely resembles the respective MHD run with $\Re = 500$. This indicates that the $\Mach = 2$ HPIC run has a smaller viscosity in comparison to that of $\Mach = 0.2$ run. These findings are consistent with our earlier visual comparison of the velocity features in Figs.~\ref{fig:slices_subsonic} and~\ref{fig:slices_supersonic}.

\section{Effective viscosity and kinetic Reynolds number of weakly- collisional plasmas} \label{sec:Re_scaling}
Here we first discuss the MHD scaling relations for the viscous and magnetic dissipation scales in \Sec{sec:Re_MHD_scaling}, and then apply these to the HPIC simulations in \Sec{sec:Re_HPIC_scaling} to infer $\Re, k_{\nu},$ and $\Pm$ for weakly-collisional plasma.

\subsection{MHD scaling relations for viscous and resistive scales} \label{sec:Re_MHD_scaling}
The viscous dissipation scale, $k_{\nu} = A\,k_{\rm turb} \Re^{\alpha}$, is related to the kinetic Reynolds number, where $k_{\rm turb}$ is the turbulence driving scale and $A$ is a dimensionless scaling coefficient. Furthermore, the ratio of the magnetic and viscous dissipation scale, $k_{\eta}/k_{\nu} = B\,\Pm^{\beta}$, can be expressed as a function of the magnetic Prandtl number \citep{KrielEtAl2022, Kriel+2023}, $B$ being another dimensionless scaling coefficient. MHD studies have investigated these scaling relations in the turbulent ($\Re > 100$) and viscous ($\Re < 100$) regimes separately \citep{Kriel+2023}. The value of the coefficients, $A$ and $B$, and the scaling exponents, $\alpha$ and $\beta$, in different $\Re$ regimes, are discussed in detail in Sec.~ 3.5 of \citet{Kriel+2023}. In this work, we provide a combined model across all $\Re$ values and refit the subsonic ($\Mach = 0.3$) and transonic ($\Mach=1$) simulations from \citet{Kriel+2023} with the present MHD runs, using a smoothly broken power-law function (motivated by the trends in the numerical results),
\begin{equation} \label{eqn:broken_power_law}
    \mathcal{BP}(q) = C \left( \frac{q}{q_{\rm b}} \right)^{\alpha} \Biggl\{ \frac{1}{2} \left[ 1 + \left( \frac{q}{q_{\rm b}} \right)^{1/\Delta} \right] \Biggl\}^{(\beta - \alpha)\Delta},
\end{equation}
where $C=\mathcal{BP}(q_{\rm b})$ is the value of the model at the transition or break point $q_{\rm b}$, $\alpha$ is the scaling exponent when $q \ll q_{\rm b}$, $\beta$ is the scaling exponent when $q \gg q_{\rm b}$, and $\Delta$ controls how smoothly the transition of the scaling exponent from $\alpha$ to $\beta$ occurs. We perform the fits for $k_{\nu}/k_{\rm turb} = \mathcal{BP}(\Re)$ and $k_{\eta}/k_{\nu} = \mathcal{BP}(\Pm)$ using the Markov chain Monte Carlo method.

\begin{table}
\centering
\caption{Fit values for parameters in \Eq{eqn:broken_power_law} shown in \Fig{fig:MHD_fits}.}
\setlength{\tabcolsep}{5pt}
\begin{tabular}{lccccc}
\hline
 & $C$ & $q_{\rm b}$ & $\alpha$ & $\beta$ & $\Delta$\\
\hline
$\mathcal{BP}(\Re)$ & 4.5$\pm$1.0 & 105$\pm$18 & 0.39$\pm$0.02 & 3/4 (fixed) & 1.8$\pm$0.2 \\
\\
$\mathcal{BP}(\Pm)$ & 1.8$\pm$0.3 & 18$\pm$4 & 1/2 (fixed) & 0.36$\pm$0.02 & 0.25$\pm$0.03 \\
\hline
\end{tabular}
\label{table:fits_MHD}
\end{table}

\begin{figure}
\begin{center}
\def\arraystretch{0}
\setlength{\tabcolsep}{0pt}
\begin{tabular}{r}
\includegraphics[width=0.98\linewidth]{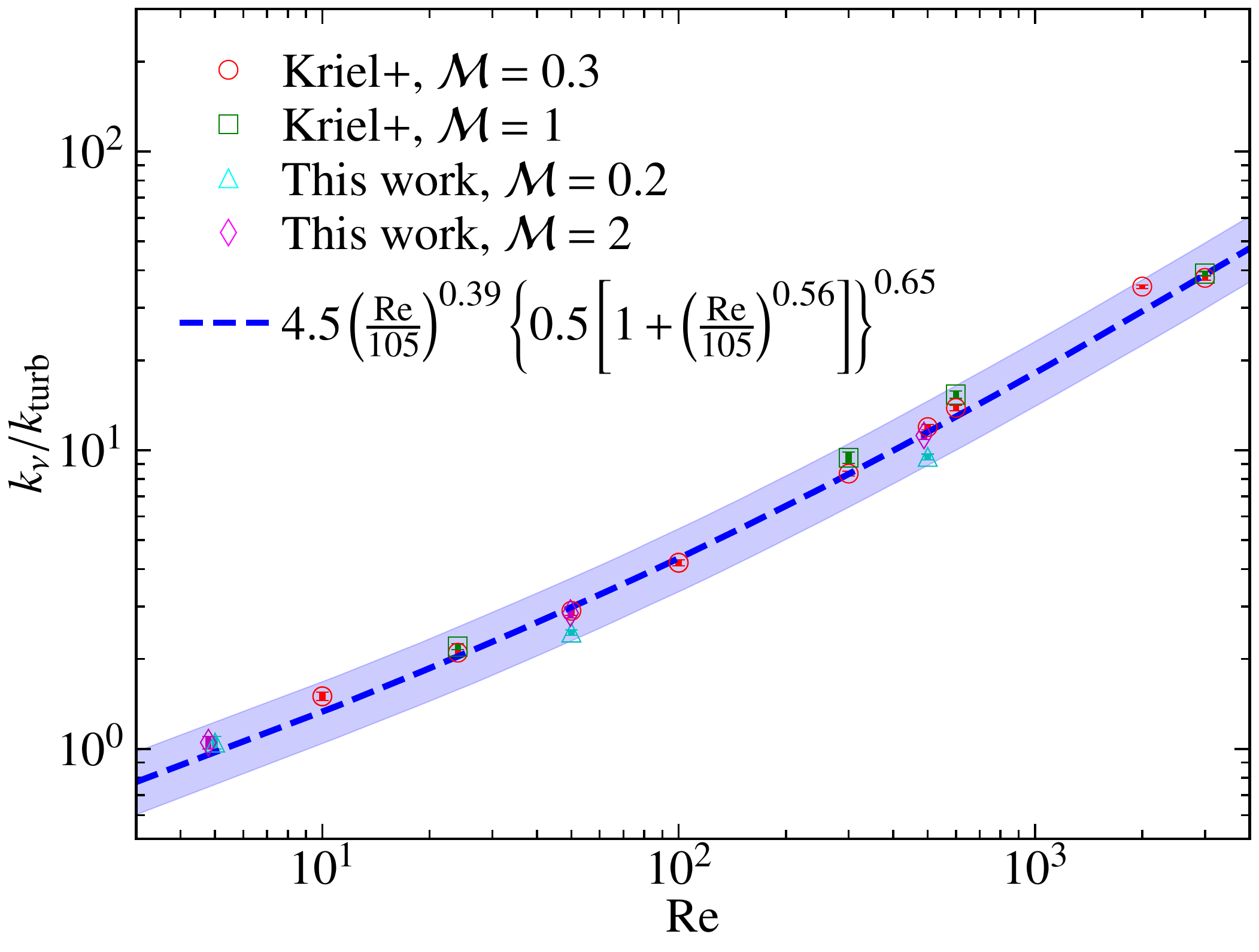} \\
\includegraphics[width=1.0\linewidth]{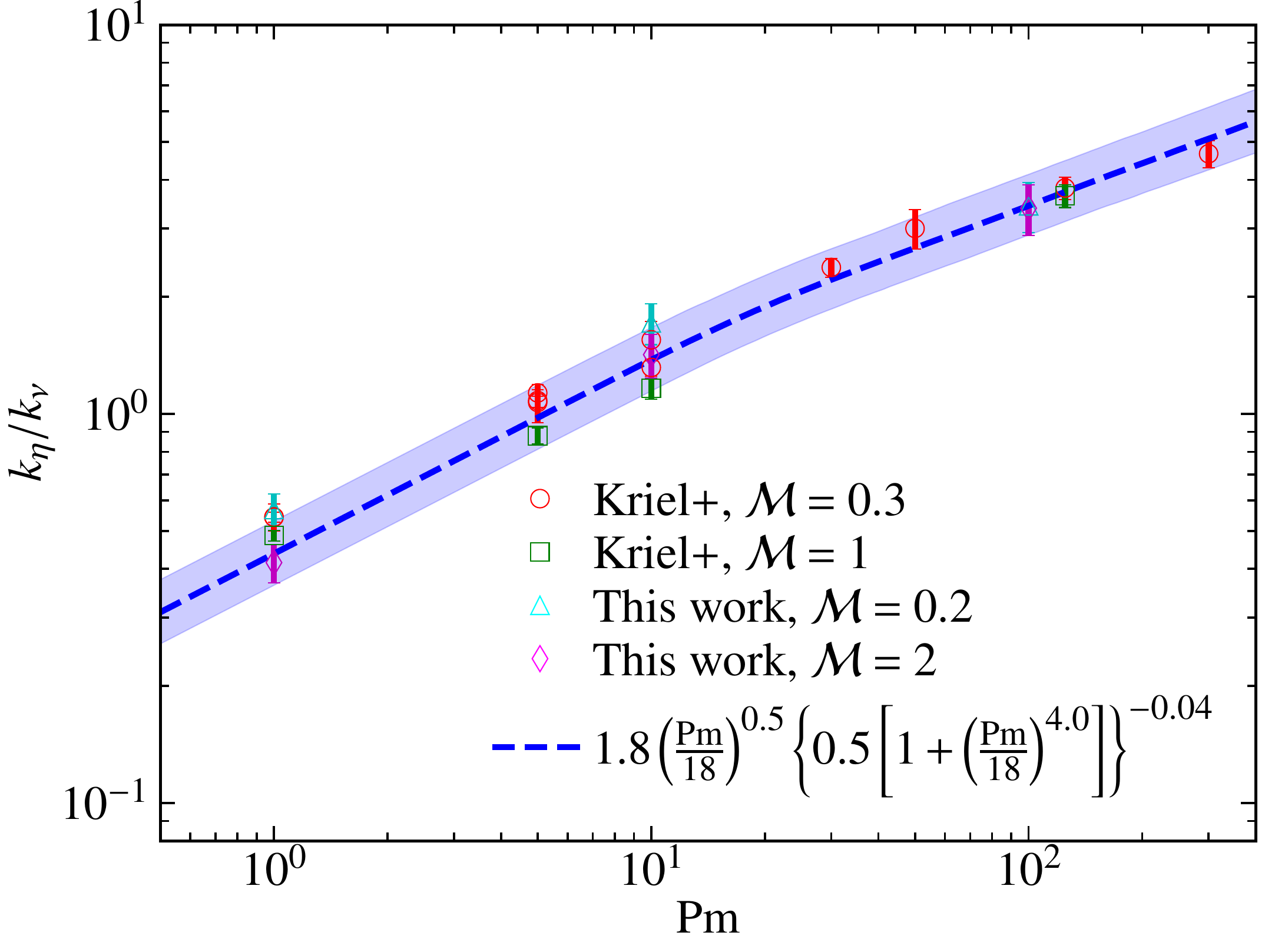}
\end{tabular}
\end{center}
\caption{Viscous dissipation wavenumber, $k_{\nu}$, normalized by the turbulent driving scale, $k_{\rm turb}$, as a function of the kinetic Reynolds number, $\Re$ (top panel), and ratio of magnetic to viscous dissipation wavenumber, $k_{\eta}/k_{\nu}$, as a function of the magnetic Prandtl number, $\Pm$ (bottom panel). We show MHD simulations with $\Mach = 0.3$ (circles) and 1 (squares) from \citet{Kriel+2023}, along with the $\Mach = 0.2$ (triangles) and 2 (diamonds) runs of this work. The dashed lines represent the median value of the fits using \Eq{eqn:broken_power_law}. The shaded region corresponds to the $16^{\rm th}$ (lower) and $84^{\rm th}$ (upper) percentile values. We find that all runs, i.e., subsonic and transonic runs up to $\mathcal{M}=2$ as investigated here, are consistent with the same scaling relations. The fitted relations and associated parameter values are listed in \Tab{table:fits_MHD}.}
\label{fig:MHD_fits}
\end{figure}

The parameters are determined by fitting the data from MHD simulations with varying Mach numbers: $\Mach = 0.2$ and 2, discussed in this work, and $\Mach = 0.3$ and 1 from \citet{Kriel+2023}. The results are shown in \Fig{fig:MHD_fits} and the fit parameters are reported in \Tab{table:fits_MHD}. We find that the scaling relations describe the MHD data reasonably well for the subsonic and transonic simulations up to $\mathcal{M} = 2$. Only for $\mathcal{M}\gtrsim5$ (not shown here), the relations change as one enters deeper into the supersonic regime, investigated in detail in Sec.~ 3.5 of \citet{Kriel+2023}.

Using $k_{\nu} = k_{\rm turb} \mathcal{BP}(\Re)$ in the expression $k_{\eta} = k_{\nu} \mathcal{BP}(\Pm)$, we can find $\Re$ from $\Rm$ (which is set via $\eta$ for both the HPIC and MHD simulations) and $k_{\eta}$ (which is determined from the peak of the current spectrum). Once $\Re$ is inferred, $k_{\nu}$ and $\Pm$ can be further estimated using \Eq{eqn:broken_power_law} and \Eq{eqn:Pm}, respectively. As a cross-check, this method is first applied on the MHD simulations themselves, with the results tabulated in \Tab{table:sims_MHD_Re}, providing a reasonable recovery of $\Re$ and $\Pm$ in all cases, as expected, since the relations were calibrated (fitted) with those data \citep[in addition to the data from][]{Kriel+2023}. Furthermore, this analysis also helps us understand the level of uncertainties in these relations.

\begin{table}
\centering
\caption{The inferred kinetic Reynolds number, $\Re_{\rm inferred}$, the viscous dissipation scale, $(k_{\nu})_{\rm inferred}$, and the magnetic Prandtl number, $\Pm_{\rm inferred}$, for the $\Mach = 0.2$ and $\Mach = 2$ runs with $\ngrid^{3} = 128^{3}$. The values are the median ($50^{\rm th}$ percentile) of the data. The lower and upper error bars show the $16^{\rm th} - 50^{\rm th}$ and $84^{\rm th} - 50^{\rm th}$ percentile of the data, respectively.}
\setlength{\tabcolsep}{5pt}
\begin{tabular}{lccc}
\hline
Model & $\Re_{\rm inferred}$ & $(k_{\nu})_{\rm inferred}$ & $\Pm_{\rm inferred}$\\
\hline
\texttt{MHDM0.2Rm500Re500} & $780_{\minus 340}^{\plus 560}$ & $31_{\minus 9.7}^{\plus 13}$ & $0.64_{\minus 0.28}^{\plus 0.53}$\\
\\
\texttt{MHDM0.2Rm500Re50} & $74_{\minus 24}^{\plus 220}$ & $7.4_{\minus 1.4}^{\plus 9.2}$ & $6.8_{\minus 5.0}^{\plus 4.2}$\\
\\
\texttt{MHDM0.2Rm500Re5} & $10_{\minus 5.6}^{\plus 45}$ & $2.7_{\minus 0.8}^{\plus 3.6}$ & $47_{\minus 38}^{\plus 68}$ \\
\\
\texttt{MHDM2Rm500Re500} & $490_{\minus 260}^{\plus 120}$ & $23_{\minus 8.8}^{\plus 3.6}$ & $1.1_{\minus 0.27}^{\plus 1.1}$\\
\\
\texttt{MHDM2Rm500Re50} & $41_{\minus 2.4}^{\plus 190}$ & $5.4_{\minus 0.17}^{\plus 8.8}$ & $11_{\minus 9.4}^{\plus 2.2}$\\
\\
\texttt{MHDM2Rm500Re5} & $10_{\minus 5.5}^{\plus 65}$ &  $2.7_{\minus 0.8}^{\plus 4.8}$ & $48_{\minus 42}^{\plus 59}$ \\
 \hline
\end{tabular}
\label{table:sims_MHD_Re}
\end{table}

\subsection{Application to HPIC simulations and the inferred $\Re, k_{\nu},$ and $\Pm$}
\label{sec:Re_HPIC_scaling}
We now apply the MHD scaling relations from above to the HPIC runs. In contrast to the MHD case, for weakly- collisional plasmas, the $\Re$ (and thus $k_{\nu}$) is a result of wave-particle interactions, and therefore not known or set a priori. Using the previously discussed MHD scaling relations, however, we can infer $\Re$ (and thus $k_{\nu}$, and $\Pm$) for the HPIC runs, assuming the MHD relations remain valid in weakly-collisional plasmas. The inferred value $\Re_{\rm inferred}$ is then used in \Eq{eqn:broken_power_law} and \Eq{eqn:Pm} to determine the viscous dissipation scale, $(k_{\nu})_{\rm inferred}$, and the magnetic Prandtl number, $\Pm_{\rm inferred}$. The values of these plasma parameters are summarised in \Tab{table:sims_Re} for all HPIC runs.

\begin{table}
\centering
\caption{Same as \Tab{table:sims_MHD_Re}, but for the HPIC simulations.}
\setlength{\tabcolsep}{5pt}
\begin{tabular}{lccc}
\hline
Model & $\Re_{\rm inferred}$ & $(k_{\nu})_{\rm inferred}$ & $\Pm_{\rm inferred}$\\
\hline
\texttt{HPICM0.2Rm500} & $480_{\minus 250}^{\plus 170}$ & $23_{\minus 8.5}^{\plus 4.9}$ & $1.1_{\minus 0.38}^{\plus 1.1}$\\
\\
\texttt{HPICM2Rm500} & $690_{\minus 360}^{\plus 360}$ & $29_{\minus 11}^{\plus 9.2}$ & $0.72_{\minus 0.29}^{\plus 0.71}$\\
\\
\texttt{HPICM0.2Rm500Nppc50} & $470_{\minus 230}^{\plus 300}$ & $22_{\minus 7.6}^{\plus 8.4}$ & $1.1_{\minus 0.47}^{\plus 0.98}$\\
\\
\texttt{HPICM0.2Rm500Nppc200} & $480_{\minus 250}^{\plus 200}$ & $23_{\minus 8.3}^{\plus 5.7}$ & $1.1_{\minus 0.41}^{\plus 1.0}$\\
\\
\texttt{HPICM2Rm500Nppc50} & $670_{\minus 340}^{\plus 410}$ & $28_{\minus 10}^{\plus 11}$ & $0.73_{\minus 0.31}^{\plus 0.69}$\\
\\
\texttt{HPICM2Rm500Nppc200} & $690_{\minus 340}^{\plus 440}$ & $29_{\minus 10}^{\plus 11}$ & $0.72_{\minus 0.31}^{\plus 0.62}$ \\
\\
\texttt{HPICM0.2Rm500Ngrid64} & $190_{\minus 73}^{\plus 310}$ &  $13_{\minus 3.1}^{\plus 10}$& $2.5_{\minus 1.6}^{\plus 1.8}$\\
\\
\texttt{HPICM0.2Rm500Ngrid256} & $440_{\minus 240}^{\plus 180}$ & $21_{\minus 8.1}^{\plus 5.2}$ & $1.2_{\minus 0.45}^{\plus 1.3}$\\
\\
\texttt{HPICM2Rm500Ngrid64} & $230_{\minus 100}^{\plus 330}$ & $14_{\minus 4.2}^{\plus 11}$ & $2.0_{\minus 1.2}^{\plus 1.6}$\\
\\
\texttt{HPICM2Rm500Ngrid256} & $770_{\minus 320}^{\plus 820}$ & $31_{\minus 9.2}^{\plus 19}$ & $0.63_{\minus 0.32}^{\plus 0.49}$\\
 \hline
\end{tabular}
\label{table:sims_Re}
\end{table}

For the standard $\Mach=0.2$ HPIC run discussed in this work, the inferred kinetic Reynolds number, $\Re_{\rm inferred} = 480_{\minus 250}^{\plus 170}$. This corresponds to $(k_{\nu})_{\rm inferred} = 23_{\minus8.5}^{\plus4.9}$ and $\Pm_{\rm inferred} = 1.1_{\minus0.38}^{\plus1.1}$. In the standard $\Mach=2$ HPIC run, the inferred values are $\Re_{\rm inferred} = 690_{\minus 360}^{\plus 360}$ and $(k_{\nu})_{\rm inferred} = 29_{\minus 11}^{\plus 9.2}$, which are somewhat larger than those in the subsonic regime. The corresponding $\Pm_{\rm inferred} = 0.72_{\minus 0.29}^{\plus 0.71}$ is somewhat lower compared to the subsonic case. While varying the number of particles per cell does not significantly affect these values, increasing the grid resolution raises the viscous dissipation scale and therefore the $\Re$ somewhat. This is because the magnetic dissipation scale increases marginally with grid resolution and with $\ngrid^{3}=64^{3}$, $\Rm = 500$ may not be resolved \citep{Malvadi&Federrath2023}.

A caveat of this approach is that the scaling relations used here are derived from numerical simulations of the MHD turbulent dynamo, which may not be directly applicable to the weakly-collisional and collisionless regime. Additionally, this approach assumes that the plasma viscosity is isotropic in space and remains unchanged in the kinematic regime of the dynamo. These assumptions may not hold true for weakly-collisional plasma. Nevertheless, the approach seems to provide reasonable estimates of Re, which are at least in qualitative agreement with other flow characteristics such as the basic structure, dynamics, and statistics of the plasma (c.f., Figs~\ref{fig:slices_subsonic} and \ref{fig:slices_supersonic} and following).

\section{Conclusions}
\label{sec:conclusions}
We study the properties of the weakly-collisional hybrid particle-in-cell (HPIC) turbulent dynamo in the subsonic ($\Mach = 0.2$) and supersonic ($\Mach = 2$) regimes in the kinematic phase using the HPIC code \ahkash and compare its properties to the collisional MHD turbulent dynamo. We use the same turbulence driving, magnetic Reynolds number, $\Rm = 500$, and initial magnetic to kinetic energy ratio, $\Einit = 10^{-10}$, for all the simulations (HPIC and MHD). For the HPIC simulations, a cooling method is used to maintain isothermal conditions and the initial ratio of the Larmor radius to the box size, $\initmagnetisation = 100$. The MHD runs have varying kinetic Reynolds and magnetic Prandtl number, $\Re = 500\;(\Pm = 1)$, $\Re = 50\;(\Pm = 10)$ and $\Re = 5\;(\Pm = 100)$. 

We compare the spatial structure, the growth rate of magnetic energy, the probability density functions (PDFs), and the power spectra of the HPIC and MHD simulations. We find that the density, velocity, and magnetic field structure of the HPIC runs is visually the closest to the MHD simulations with $\Re\sim50-500$, as discussed in \Sec{sec:slice_plots}. The growth rate of magnetic energy in HPIC is similar to that of MHD with $\Pm = 1-100$ in the subsonic regime of turbulence, whereas in the supersonic regime, it resembles that of MHD with $\Pm = 1-10$, as shown in \Sec{sec:time_evol}.

We fit the density PDFs with a lognormal and a Castaing-Hopkins distribution in \Sec{sec:dens_pdfs} and find that the lognormal distribution is a suitable model for all runs. For the velocity PDF, discussed in \Sec{sec:vels_pdfs}, a Gaussian distribution with mean $\sim 0$ and standard deviation $\sim1/\sqrt{3}$ fits well, as expected. The magnetic field PDFs are fit with a Cauchy-normal distribution, showing that both MHD and HPIC turbulent dynamo exhibit highly intermittent magnetic fields (see \Sec{sec:mags_pdfs}).

The magnetic energy power spectra of HPIC are similar to those of MHD with $\Re \sim 50-500$, as shown in \Sec{sec:mags_spectra}. We find that the dissipation scale of the magnetic energy, $k_{\eta}$, of the HPIC runs, is $9.5 \pm 1.0$ and $9.9 \pm 1.1$ in the subsonic and supersonic regime, respectively. The magnetic dissipation scale for the subsonic HPIC run is comparable to that of the MHD simulation with $\Re = 50$–$500$, while for the supersonic HPIC run, it is similar to the MHD run with $\Re = 500$ (see \Sec{sec:current_spectra}). The turbulent kinetic energy power spectrum in the subsonic HPIC run is similar to the respective MHD runs with $\Re \sim 50-500$, while in the supersonic case, it is similar to the MHD run with $\Re \sim 500$, as discussed in \Sec{sec:vels_spectra}. Applying scaling relations determined from studies of the MHD turbulent dynamo in \Sec{sec:Re_MHD_scaling}, we find $\Re = 480_{\minus250}^{\plus170}$ for subsonic HPIC, while supersonic HPIC has $\Re = 690_{\minus360}^{\plus360}$, as discussed in \Sec{sec:Re_HPIC_scaling}. The corresponding viscous dissipation scale, $k_{\nu} = 23_{\minus8.5}^{\plus4.9}$ and $29_{\minus11}^{\plus9.2}$ in the subsonic and supersonic HPIC runs, respectively.

Overall, we find that the turbulent dynamo exhibits similar physical properties in weakly-collisional (HPIC) and collisional (MHD) plasmas. For subsonic turbulence, the present HPIC simulations share similarities with MHD at $\Re\sim50-500$, while in the supersonic regime, HPIC is similar to MHD at $\Re\sim500$. To the best of our knowledge, this is the first time that the weakly-collisional plasma dynamo has been studied in the supersonic regime of turbulence, which may be relevant for weakly-collisional shocks (e.g., potentially in supernovae or the supersonic regions of the solar wind), and the compressible regions of the galaxy intracluster medium (ICM).

\section*{Acknowledgements}
R.~A.~C.~ acknowledges financial support from the Australian Government via the Australian Government Research Training Program Fee-Offset Scholarship. R.~A.~C.~also acknowledges that this work was supported by an NCI HPC-AI Talent Program 2023 Scholarship (project gp08), with computational resources provided by NCI Australia, an NCRIS-enabled capability supported by the Australian Government. C.~F.~acknowledges funding provided by the Australian Research Council (Discovery Project grants~DP230102280 and~DP250101526), and the Australia-Germany Joint Research Cooperation Scheme (UA-DAAD). A.~S.~acknowledges support from the Australian Research Council's Discovery Early Career Researcher Award (DECRA, project~DE250100003). We further acknowledge high-performance computing resources provided by the Leibniz Rechenzentrum and the Gauss Centre for Supercomputing (grants~pr32lo, pr48pi, pn76ga and GCS Large-scale project~10391), the Australian National Computational Infrastructure (grant~ek9) and the Pawsey Supercomputing Centre (project~pawsey0810) in the framework of the National Computational Merit Allocation Scheme and the ANU Merit Allocation Scheme. The simulation software, \texttt{FLASH}, was in part developed by the Flash Centre for Computational Science at the University of Chicago and the Department of Physics and Astronomy at the University of Rochester.

\section*{Data Availability}
The simulation data (5~TB) will be shared on reasonable request to the corresponding author.


\bibliographystyle{mnras}
\bibliography{Ref_files_Radhika.bib} 




\appendix

\section{Particle and grid resolution convergence tests for hybrid PIC simulations}
\label{app:hybridPIC_conv}

\begin{figure}
    \includegraphics[scale = 0.3 ]{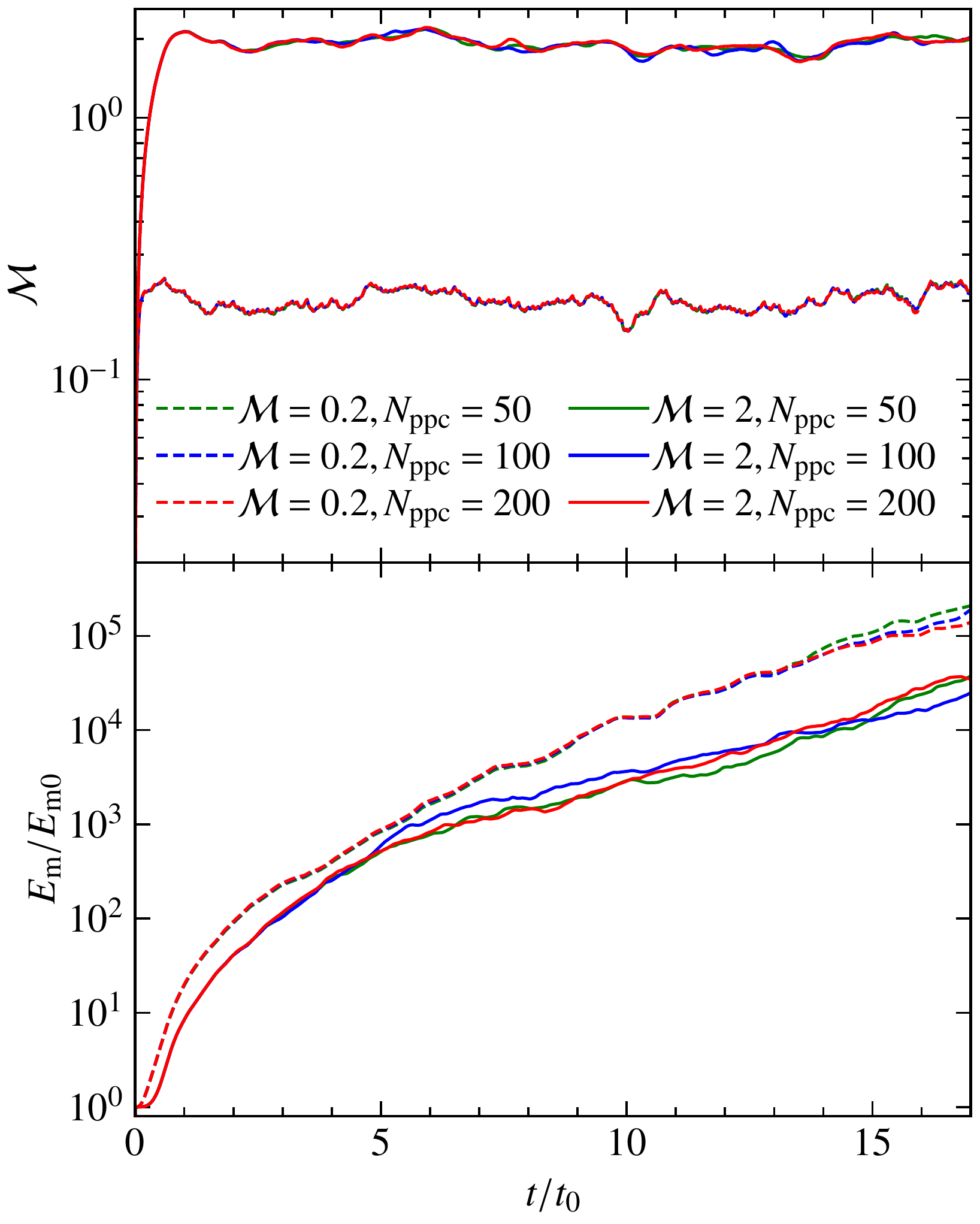}
    \caption{Same as \Fig{fig:time_evol}, but for subsonic and supersonic weakly-collisional turbulent dynamo simulations with fixed grid resolution $\ngrid^{3} = 128^3$ and varying $\nppc = 50, 100$ and 200 particles per cell. The Mach number and magnetic energy growth rate are tabulated in \Tab{table:sims}. The numerical simulations show convergence with $\nppc=100$.}
    \label{fig:ppc_test}
\end{figure}

To show the convergence of our $\Mach=0.2$ and $\Mach=2$ simulations, we repeat our numerical experiments with varying grid resolutions and number of particles per cell. The time evolution of the HPIC runs with a fixed grid resolution ($\ngrid^3 = 128^3$) and varying particle resolution of $\nppc=50$, $100$, and $200$ particles per cell is shown in \Fig{fig:ppc_test}. We do not find any significant difference in the time evolution of the dynamo as the number of particles per cell is changed.

\begin{figure}
    \includegraphics[scale = 0.3]{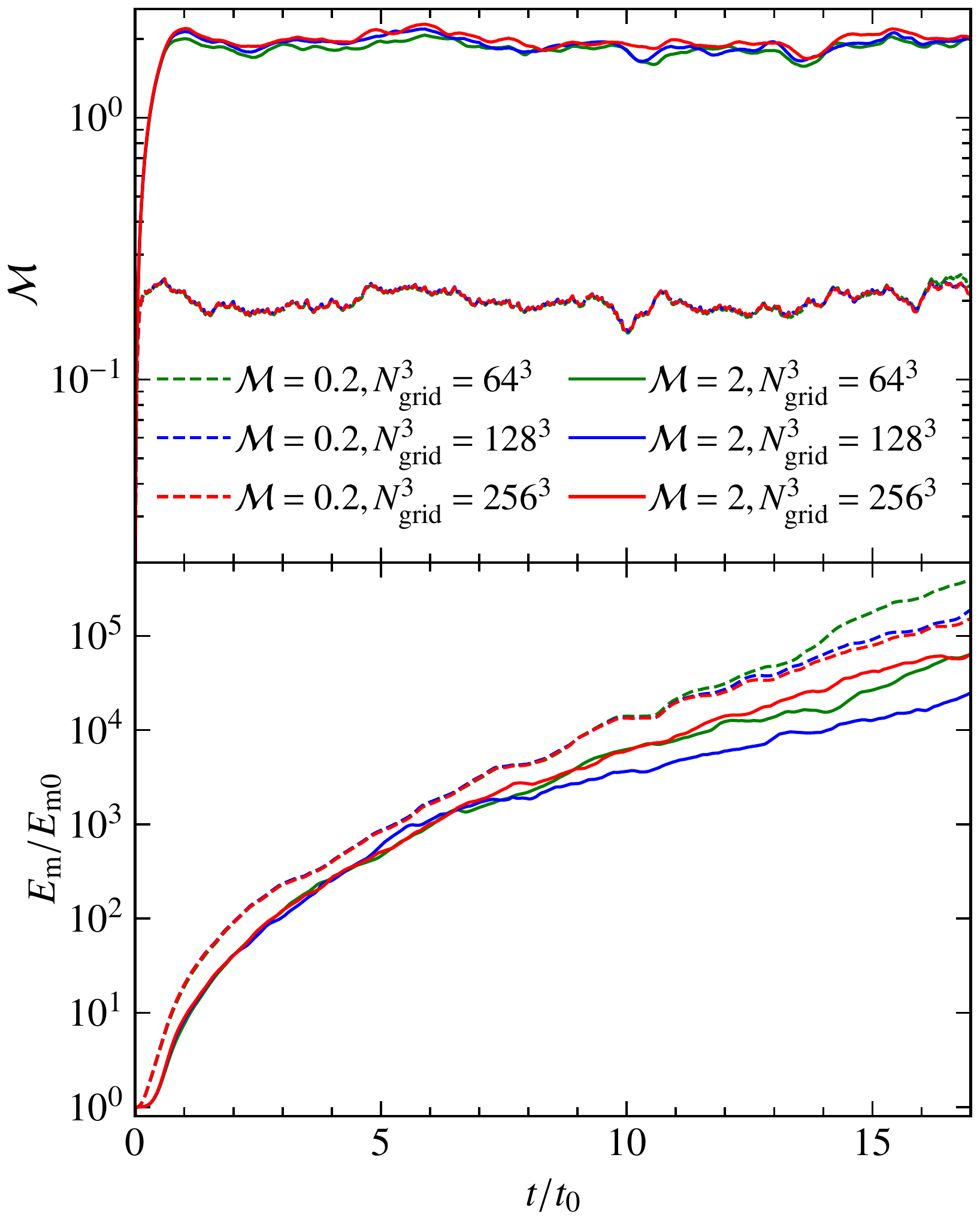}
    \caption{Same as \Fig{fig:ppc_test}, but for subsonic and supersonic weakly-collisional turbulent dynamo simulations with fixed $\nppc = 100$ particles per cell and varying grid resolution of $\ngrid^3 = 64^3, 128^3$ and $256^3$. The Mach number and growth rate are converged with $\ngrid^3=128^3$ in both the subsonic and supersonic regimes.}
    \label{fig:grid_test}
\end{figure}

The time evolution of the HPIC simulations with a fixed number of particles-per-cell ($\nppc = 100$) and varying grid resolution of $64^{3}$, $128^{3}$, and $256^{3}$ is shown in \Fig{fig:grid_test}. We find that the Mach number and magnetic energy growth rate are converged with $\ngrid^3=128^3$ (see \Tab{table:sims}).

\section{Grid resolution convergence test for MHD simulations}
\label{app:MHD_conv}

\begin{figure}
    \includegraphics[scale = 0.3]{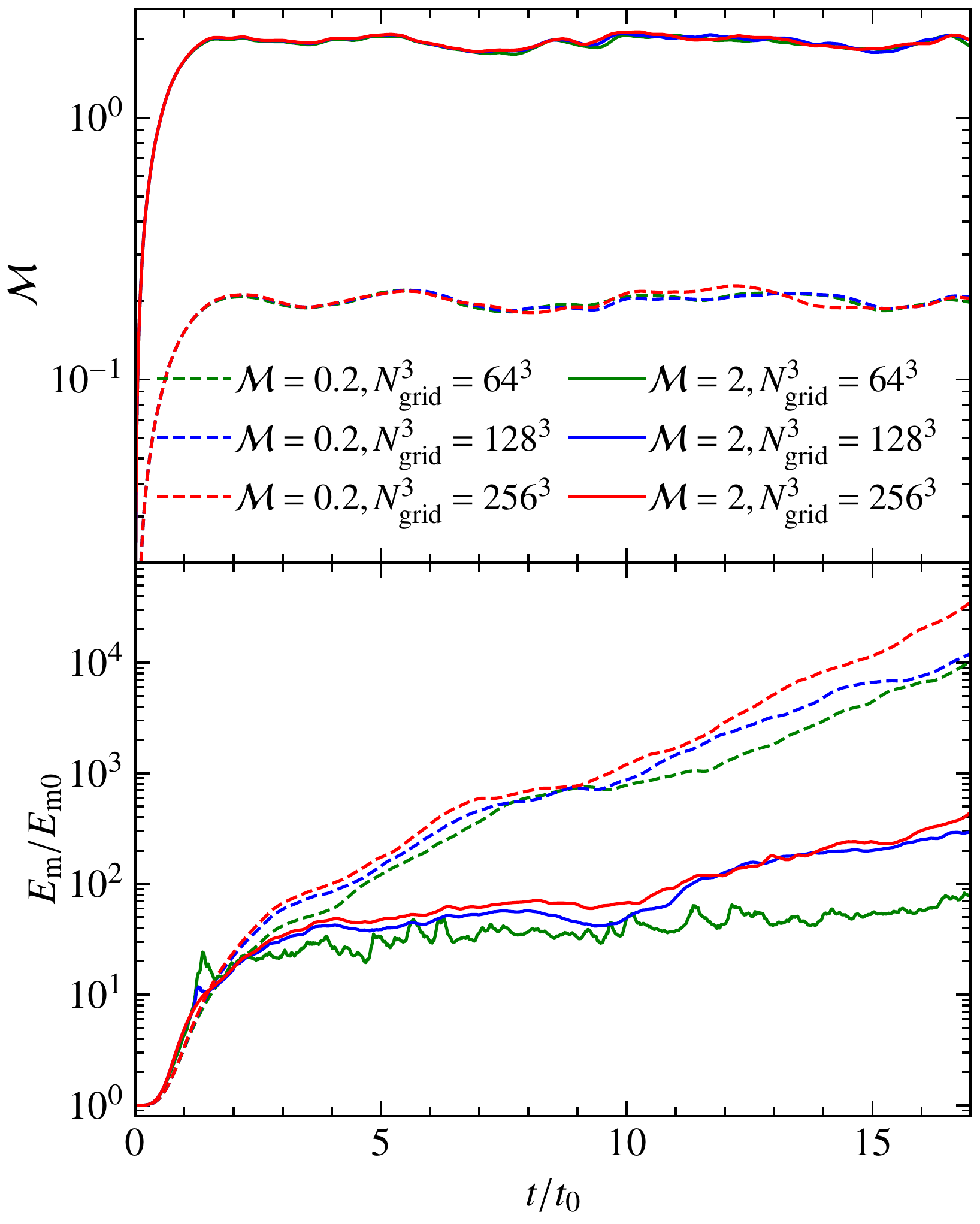}
    \caption{Same as \Fig{fig:grid_test}, but for subsonic and supersonic MHD turbulent dynamo simulations with kinetic Reynolds number, $\Re=500$. The numerical simulations show convergence with $\ngrid^3=128^3$.}
    \label{fig:grid_test_MHD}
\end{figure}

To show the convergence of our subsonic and supersonic MHD simulations, we repeat our numerical experiments with varying grid resolution, $\ngrid^3 = 64^{3}$, $128^{3}$, and $256^{3}$. We find that the Mach number and magnetic energy growth rate are converged with $\ngrid^3=128^3$ for both the $\Mach=0.2$ and $\Mach=2$ MHD simulations.


\bsp	
\label{lastpage}
\end{document}